\documentclass[conference]{IEEEtran}
\IEEEoverridecommandlockouts
\usepackage{
amsmath,cite,bm,amsfonts,amssymb,
graphicx,color,mathtools,setspace,
comment,multirow,xfrac,epstopdf,
footnote,multirow,lscape,float,
caption,subcaption,amsthm,alltt,placeins,amsthm}
\usepackage[ruled,vlined]{algorithm2e}

\newcommand{\myincludegraphics}{\includegraphics[trim=0cm 0cm 0cm 0.7cm, clip=true, width=\columnwidth]}

\newcommand{\myVM}[3]{\mathbf{#1}_{\mathrm{#2}}^{#3}} 
\newcommand{\myVMIndex}[4]{\mathbf{#1}_{\mathrm{#2},#3}^{#4}} 
\newcommand{\Norm}[3]{\left\lVert #1 \right\rVert_{#2}^{#3}} 
\newcommand{\tr}[3]{\mathrm{tr}\left\{\mathbf{#1}_{\mathrm{#2}}^{#3}\right\}}
\newcommand{\eq}[1]{(#1)}
\newcommand{\Diag}[1]{\mathrm{diag}\left\{#1\right\}}
\newcommand{\Det}[1]{\left\lvert #1 \right\rvert}
\newcommand{\Abs}[2]{\left\lvert #1 \right\rvert^{#2}}
\newcommand{\Exp}[1]{\mathbb{E}\left\{ #1 \right\}}

\newcommand{\kth}[1]{#1^{\mathrm{th}}}

\newcommand{\betabr}[0]{\beta_{\mathrm{br}}}

\newcommand{\etad}[0]{\eta_{\mathrm{d}}}
\newcommand{\zetad}[0]{\zeta_{\mathrm{d}}}
\newcommand{\etaru}[0]{\eta_{\mathrm{ru}}}
\newcommand{\zetaru}[0]{\zeta_{\mathrm{ru}}}
\newcommand{\etabr}[0]{\eta_{\mathrm{br}}}
\newcommand{\zetabr}[0]{\zeta_{\mathrm{br}}}

\newcommand{\kd}[0]{\kappa_{\mathrm{d}}}
\newcommand{\kru}[0]{\kappa_{\mathrm{ru}}}
\newcommand{\kbr}[0]{\kappa_{\mathrm{br}}}

\newcommand{\Hd}[1]{\myVM{H}{d}{#1}}
\newcommand{\Hru}[1]{\myVM{H}{ru}{#1}}
\newcommand{\Hbr}[1]{\myVM{H}{br}{#1}}
\newcommand{\Ubr}[1]{\myVM{U}{}{#1}}

\newcommand{\wwr}[1]{\myVM{w}{}{#1}}

\newcommand{\ab}[1]{\myVM{a}{b}{#1}}
\newcommand{\ar}[1]{\myVM{a}{r}{#1}}

\newcommand{\argmax}[1]{\underset{#1}{\text{argmax}}}

\newcommand{\maxover}[1]{\underset{#1}{\text{max}}}

\begin{document}

\bstctlcite{IEEEexample:BSTcontrol} 

\title{Efficient Optimization Techniques for RIS-aided Wireless Systems}

\author{
	\IEEEauthorblockN{%
		Ikram Singh\IEEEauthorrefmark{1}, %
		Peter J. Smith\IEEEauthorrefmark{2}, %
		Pawel A. Dmochowski\IEEEauthorrefmark{1}, %
		Rua Murray\IEEEauthorrefmark{3}}
	\IEEEauthorblockA{\IEEEauthorrefmark{1}%
		School of Engineering and Computer Science, Victoria University of Wellington, Wellington, New Zealand}
	\IEEEauthorblockA{\IEEEauthorrefmark{2}%
		School of Mathematics and Statistics, Victoria University of Wellington, Wellington, New Zealand}
	\IEEEauthorblockA{\IEEEauthorrefmark{3}%
	    School of Mathematics and Statistics, University of Canterbury,
	    Christchurch, New Zealand}
	\IEEEauthorblockA{email:%
		~\{ikram.singh, peter.smith, pawel.dmochowski\}@ecs.vuw.ac.nz,
		rua.murray@canterbury.ac.nz
	}%
}

\maketitle
\begin{abstract}
The objective of this paper is to develop simple techniques to enhance the performance of multi-user RIS aided wireless systems. Specifically, we develop a novel technique called \textit{channel separation} which provides a better understanding of how the RIS phases affect the uplink sum rate and sum rates for ZF and MMSE linear receivers. Leveraging channel separation, we propose a simple iterative algorithm to improve the uplink sum rate and the sum rates of ZF and MMSE linear receivers when discrete RIS phases are considered. For continuous RIS phases, we derive simple closed form solutions to enhance the uplink sum rate and reduce the total mean square error of the MMSE combiner. The latter metric is a tractable alternative to maximizing sum rates for ZF and MMSE. Numerical simulations are performed for all optimization techniques and the effectiveness of each technique is compared to a full numerical optimization procedure, namely an interior point (IP) algorithm.
%
\end{abstract}
\IEEEpeerreviewmaketitle
%
%
\section{Introduction}
Reconfigurable Intelligent Surfaces (RIS) are an important technology for future wireless communications, due to their ability to manipulate the channel between users (UEs) and a base station (BS). Assuming that channel state information (CSI) is known, it is possible to intelligently configure the RIS phases to optimize metrics such as system sum-rate, energy efficiency or secrecy rate. However, it has become apparent that the unit modulus constraint at the RIS, where only the phases and not the amplitudes of reflected signals can be controlled, makes any optimization of system metrics extremely difficult. Furthermore, practical scenarios where the RIS phases are selected from a discrete set further complicates the optimization problems. For this reason, much of the literature has focused on complex, numerical approaches which give bounds or yield high high performance with relatively high complexity methods.

Maximizing the sum-rate for multi-user (MU) systems with a single RIS is considered in 
\cite{9110889,9117136,9148766,8849960,9013288,9203956,9238961,9286726,9316283,9320618}. Specifically, \cite{9110889} develops a hybrid beamforming scheme where digital beamforming is performed at the BS and analog beamforming is used at the RIS for discrete RIS phases. This is achieved through an iterative algorithm which utilises the branch-and-bound method. Results show that the system can achieve a good sum-rate performance even with low resolution RIS phases. Iterative algorithms designed to solve joint optimization problems are also proposed in \cite{9148766,9286726,9320618}. In \cite{9117136}, a sample average approximation (SAA) algorithm is designed but with continuous RIS phases. A local search method is proposed in \cite{8849960} under discrete RIS phases. In \cite{9013288}, the weighted sum rate is maximized through joint optimization of the active and passive beamforming at the BS and RIS, respectively. This is achieved through an alternating optimization method for each beamforming problem which is initially decomposed using the Lagrangian dual transform. Here, passive beamforming optimization is used for both discrete and continuous RIS phases. A similar joint optimization problem is designed in \cite{9316283} and solved through a robust beamforming design utilizing the penalty dual decomposition (PDD) algorithm.

Evidently, iterative algorithms have proven to be very useful tools in optimizing the sum-rate of RIS-aided wireless systems. Furthermore, the majority of the literature maximizes the sum-rate via a joint optimization of the beamforming vector at the BS and the RIS phases. However, there is a clear gap in the literature around efficient optimization techniques for the sum-rate of existing linear processors, such as zero-forcing (ZF) and minimum-mean-square-error (MMSE) receivers.

Hence, in this paper, we make the following contributions:
\begin{itemize}
    \item We introduce channel separation, a very powerful tool which enables a wide variety of complex RIS design problems to be collapsed to and approximated by much simpler problems involving quadratic forms for which approximate optimization is possible.
    \item In particular, channel separation is used to provide RIS designs which enhance the uplink sum rate, $R_{\mathrm{sum}}$, the sum rate for a ZF receiver, $R_{\mathrm{ZF}}$, and the sum rate for an MMSE receiver, $R_{\mathrm{MMSE}}$.
    \item The channel separation approach also leads to design problems which can handle both low-bit and high-bit/continuous phase operation at the RIS.
    \item With discrete RIS phases at the RIS, $R_{\mathrm{sum}},R_{\mathrm{ZF}}$ and $R_{\mathrm{MMSE}}$ can be enhanced using an alternating optimization based search algorithm.
    \item With continuous RIS phases at the RIS, we develop a practical solution to enhance $R_{\mathrm{sum}}$. For $R_{\mathrm{ZF}}$ and $R_{\mathrm{MMSE}}$, we note that these sum rate metrics are very complex non-linear functions of the RIS phases for which closed form designs are very challenging. Hence, we focus on the total mean squared error, $\mathrm{MSE}_{\mathrm{Tot}}$, of the MMSE receiver as an alternative metric due to its relative simplicity and strong link to receiver performance. Further, we develop a low complexity solution to enhance $\mathrm{MSE}_{\mathrm{Tot}}$.
    \item Simulation results with general ray-based channel models are conducted to support our optimization techniques.
\end{itemize}
These contributions extend considerably the earlier conference paper  \cite{TightBounds}  which applied channel separation in the continuous case only to the single metric, $R_{\mathrm{sum}}$.

\textit{Notation:} $\Norm{\cdot}{2}{}$ denotes the $\ell_{2}$ norm. The transpose, Hermitian transpose and complex conjugate operators are denoted as $(\cdot)^{T},(\cdot)^{H},(\cdot)^{*}$, respectively. The angle of a vector $\myVM{x}{}{}$ of length $N$ is defined as $\angle \myVM{x}{}{} = [ \angle x_{1},\ldots, \angle x_{N} ]^{T}$ and the exponent of a vector is defined as $e^{\myVM{x}{}{}} = [ e^{x_{1}},\ldots,  e^{x_{N}} ]^{T}$. The Kronecker product is denoted $\otimes$. $\mathcal{U}[a,b]$ denotes a uniform distribution on the interval $[a,b]$, $\mathcal{N}(\mu,\sigma^2)$ denotes a Normal distribution with mean $\mu$ and variance $\sigma^2$ and $\mathcal{L}(1/{\sigma})$ denotes a Laplacian distribution with standard deviation parameter $\sigma$. $\Abs{\mathbf{X}}{}$ denotes the determinant of a matrix $\mathbf{X}$. $\Re\{\cdot\}$ denotes the real operator.

\section{Channel and System Model}\label{Sec: Channel Model}
\begin{figure}[h]
	\centering
	\includegraphics[width=7cm]{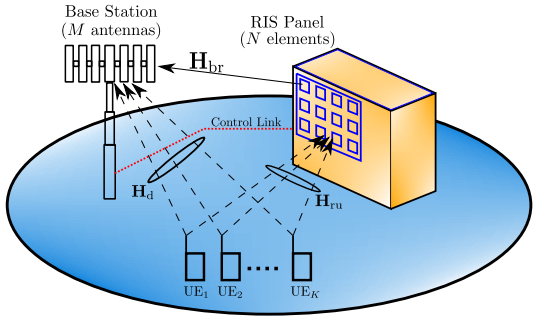}
	\caption{System model.}
	\label{Fig: System Model}
\end{figure}

As shown in Fig.~\ref{Fig: System Model}, we examine a RIS-aided wireless system where a RIS with $N$ reflective elements supports  UL transmission between $K$ single antenna UEs and a BS with $M$ antennas. 

Let $\myVM{H}{d}{} \in \mathbb{C}^{M \times K}$, $\myVM{H}{ru}{} \in \mathbb{C}^{N \times K}$, $\myVM{H}{br}{} \in \mathbb{C}^{M \times N}$ be the UE-BS, UE-RIS, RIS-BS channels, respectively. The diagonal matrix $\myVM{\Phi}{}{} \in \mathbb{C}^{N \times N}$, where $\mathbf{\Phi}_{rr} = e^{j\phi_r}$ for $r=1,2,\ldots,N$, contains the reflection coefficients for each RIS element. Given these matrices, the global UL channel is given by
\begin{equation}\label{Eq: Global Channel}
	\myVM{H}{}{} = \myVM{H}{d}{} + \myVM{H}{br}{}\mathbf{\Phi}\myVM{H}{ru}{}.
\end{equation}
In the channel model, we adopt a LOS version of the clustered, ray-based model in \cite{8812955} for $\myVM{H}{d}{}, \myVM{H}{ru}{}$:
\begin{equation}\label{Eq: Channels}
	\begin{split}
		\myVM{H}{d}{} & =  \etad \myVM{A}{d}{\mathrm{LOS}}\myVM{B}{d}{1/2} + \zetad\sum_{c=1}^{C_{\mathrm{d}}}\sum_{s=1}^{S_{\mathrm{d}}}\myVM{A}{d,c,s}{\mathrm{SC}}, \\
		\myVM{H}{ru}{} & = \etaru \myVM{A}{ru}{\mathrm{LOS}}\myVM{B}{ru}{1/2} + \zetaru\sum_{c=1}^{C_{\mathrm{ru}}}\sum_{s=1}^{S_{\mathrm{ru}}}\myVM{A}{ru,c,s}{\mathrm{SC}},
	\end{split}
\end{equation}
with 
\begin{align*}
    \etad = \sqrt{\frac{\kd}{1+\kd}}, &\quad \zetad = \sqrt{\frac{1}{1+\kd}}, \\
	\etaru = \sqrt{\frac{\kru}{1+\kru}}, &\quad \zetaru = \sqrt{\frac{1}{1+\kru}},
\end{align*}
where $C_{\mathrm{d}},C_{\mathrm{ru}}$ are the number of clusters in the UE-BS, UE-RIS channels, and $S_{\mathrm{d}},S_{\mathrm{ru}}$ are the number of sub-rays per cluster in the UE-BS and UE-RIS channels. In \eq{\ref{Eq: Channels}}, $\kd$ and $\kru$ are the equivalent of Ricean K-factors for the UE-BS and UE-RIS channels, respectively, controlling the relative power of the scattered (ray-based) components and the LOS ray. For simplicity, we assume that each user has the same K-factor, but this can easily be generalized. $\myVM{B}{d}{},\myVM{B}{ru}{}$ are diagonal matrices containing the path gains between UE-BS and UE-RIS respectively, which are modeled by distance-dependent path loss. In particular
\begin{equation}\label{Eq: Path gain matrices for Hd and Hru}
    \left( \myVM{B}{d}{} \right)_{kk} = P d_{\mathrm{d},k}^{-\gamma_{\mathrm{d}}},
    \quad
    \left( \myVM{B}{ru}{} \right)_{kk} = P d_{\mathrm{ru},k}^{-\gamma_{\mathrm{ru}}},
\end{equation}
where $d_{\mathrm{d},k}$ and $d_{\mathrm{ru},k}$ are the distances between the $k^{\mathrm{th}}$ UE and the BS and the $k^{\mathrm{th}}$ UE and the RIS respectively, $\gamma_{\mathrm{d}}$ and $\gamma_{\mathrm{ru}}$ are the pathloss exponents, $P$ is the path loss at a reference distance of 1m.
$\myVM{A}{d}{\mathrm{LOS}}$ and $\myVM{A}{ru}{\mathrm{LOS}}$ are the LOS components for the UE-BS and UE-RIS channels respectively. The $\kth{k}$ columns of the LOS components for $\Hd{}$ and $\Hru{}$ are given by
\begin{equation}\label{Eq: LOS Component for Hd and Hru}
    \myVM{a}{\mathrm{d},k}{\mathrm{LOS}} = \mathbf{a}_{\mathrm{b}}(\theta_{\mathrm{d}}^{(k)},\phi_{\mathrm{d}}^{(k)}), \
    \myVM{a}{\mathrm{ru},k}{\mathrm{LOS}} = \mathbf{a}_{\mathrm{r}}(\theta_{\mathrm{ru}}^{(k)},\phi_{\mathrm{ru}}^{(k)}),
\end{equation}
where $\theta_{\mathrm{d}}^{(k)}$, $\theta_{\mathrm{ru}}^{(k)}$ are the elevation angles of arrival (AOAs) for the $\kth{k}$ UE and $\phi_{\mathrm{d}}^{(k)}$, $\phi_{\mathrm{ru}}^{(k)}$ are the azimuth AOAs for the $\kth{k}$ UE. Note that the steering vectors at the BS, $\ab{}(\cdot,\cdot)$, and at the RIS, $\ar{}(\cdot,\cdot)$, are topology dependent. Further details are given in Sec.~\ref{Sec: Results}.
Finally $\myVM{A}{d,c,s}{\mathrm{SC}}$ and $\myVM{A}{ru,c,s}{\mathrm{SC}}$ are the scattered components due to the $s$-th subray in the $c$-th cluster which are modeled as in \cite{8812955}. The $k^{\mathrm{th}}$ columns of $\myVM{A}{d,c,s}{\mathrm{SC}}$ and $\myVM{A}{ru,c,s}{\mathrm{SC}}$ are given by the weighted steering vectors,
\begin{equation}\label{Eq: Scattered component for Hd and Hru}
\begin{split}
    \myVMIndex{a}{d,c,s}{k}{\mathrm{SC}} &= \gamma_{\mathrm{d},c,s}^{(k)}\myVM{a}{\mathrm{b}}{}(\theta_{\mathrm{d},c,s}^{(k)},\phi_{\mathrm{d},c,s}^{(k)}), \\
    \myVMIndex{a}{ru,c,s}{k}{\mathrm{SC}} &= \gamma_{\mathrm{ru},c,s}^{(k)}\myVM{a}{\mathrm{r}}{}(\theta_{\mathrm{ru},c,s}^{(k)},\phi_{\mathrm{ru},c,s}^{(k)}),
\end{split}
\end{equation}
where $\theta_{\mathrm{d},c,s}^{(k)}$, $\theta_{\mathrm{ru},c,s}^{(k)}$ are the elevation AOAs and $\phi_{\mathrm{d},c,s}^{(k)}$, $\phi_{\mathrm{ru},c,s}^{(k)}$ are the azimuth AOAs experienced by the $\kth{k}$ UE. The elevation AOAs are calculated by $\theta_{\mathrm{d},c,s}^{(k)} = \theta_{\mathrm{d},c}^{(k)} + \delta_{\mathrm{d},c,s}^{(k)}$ and $\theta_{\mathrm{ru},c,s}^{(k)} = \theta_{\mathrm{ru},c}^{(k)} + \delta_{\mathrm{ru},c,s}^{(k)}$ where $\theta_{\mathrm{d},c}^{(k)},\theta_{\mathrm{ru},c}^{(k)}$ are the central angles for the subrays in cluster $c$ and the deviations of the subrays from the central angle are $\delta_{\mathrm{d},c,s}^{(k)},\delta_{\mathrm{ru},c,s}^{(k)}$. The azimuth AOAs for each ray are $\phi_{\mathrm{d},c,s}^{(k)} = \phi_{\mathrm{d},c}^{(k)} + \Delta_{\mathrm{d},c,s}^{(k)}$ and $\phi_{\mathrm{ru},c,s}^{(k)} = \phi_{\mathrm{ru},c}^{(k)} + \Delta_{\mathrm{ru},c,s}^{(k)}$ where $\phi_{\mathrm{d},c}^{(k)},\phi_{\mathrm{ru},c}^{(k)}$ are the central angles for the subrays in cluster
$c$ and the deviations of the subrays from the central angle are $\Delta_{\mathrm{d},c,s}^{(k)}, \Delta_{\mathrm{ru},c,s}^{(k)}$. $\gamma_{\mathrm{d},c,s}^{(k)}=\beta_{\mathrm{d},c,s}^{(k)1/2}e^{j\psi_{\mathrm{d},c,s}^{(k)}}$ and $\gamma_{\mathrm{ru},c,s}^{(k)}=\beta_{\mathrm{ru},c,s}^{(k)1/2}e^{j\psi_{\mathrm{ru},c,s}^{(k)}}$ are the ray coefficients where the random phases satisfy  $\psi_{\mathrm{d},c,s}^{(k)},\psi_{\mathrm{ru},c,s}^{(k)} \sim
\mathcal{U}(0,2\pi)$ and the ray powers $\beta_{\mathrm{d},c,s}^{(k)}$ and $\beta_{\mathrm{ru},c,s}^{(k)}$ satisfy $\left(\mathbf{B}_{\mathrm{d}}\right)_{kk} =
\sum_{c=1}^{C_{\mathrm{d}}}\sum_{s=1}^{S_{\mathrm{d}}} \beta_{\mathrm{d},c,s}^{(k)}$ and
$\left(\mathbf{B}_{\mathrm{ru}}\right)_{kk} = \sum_{c=1}^{C_{\mathrm{ru}}}\sum_{s=1}^{S_{\mathrm{ru}}} \beta_{\mathrm{ru},c,s}^{(k)}$.

The majority of the results in this paper are for a pure LOS RIS-BS channel. However, we also show numerically that the results can be applied to scenarios where  $\Hbr{}$ has a smaller scattered component and a dominant LOS path. Hence, we consider the following channel models.
\subsubsection{$\Hbr{}$ is pure LOS}
\begin{equation}\label{Eq: Hbr pure LOS}
	\Hbr{} = \sqrt{\betabr}\myVM{A}{br}{\mathrm{LOS}},
\end{equation}
with
\begin{equation}\label{Eq: LOS Component for Hbr}
    \myVM{A}{br}{\mathrm{LOS}} = \mathbf{a}_{\mathrm{b}}(\theta_{\mathrm{br,A}},\phi_{\mathrm{br,A}})
    \mathbf{a}_{\mathrm{r}}^{H}(\theta_{\mathrm{br,D}},\phi_{\mathrm{br,D}}),
\end{equation}
where $\theta_{\mathrm{br,A}}$, $\phi_{\mathrm{br,A}}$ are the elevation and azimuth AOAs and $\theta_{\mathrm{br,D}}$, $\phi_{\mathrm{br,D}}$ are the elevation and azimuth angles of departure (AODs), $\betabr$ is the link gain between RIS and BS. Here, $\Hbr{}$ is rank-1 and the path gain is $\betabr = d_{br}^{-2}$, where $d_{\mathrm{br}}$ is the distance between RIS-BS.
\subsubsection{$\Hbr{}$ is dominant LOS}
\begin{equation}\label{Eq: Hbr is dominant LOS}
	\myVM{H}{br}{} = \etabr \sqrt{\betabr} \myVM{A}{br}{\mathrm{LOS}} + \zetabr\sum_{c=1}^{C_{\mathrm{br}}}\sum_{s=1}^{S_{\mathrm{br}}}\myVM{A}{br,c,s}{\mathrm{SC}},
\end{equation}
such that $\etabr >> \zetabr$, with
$$
	\etabr = \sqrt{\frac{\kbr}{1+\kbr}}, \quad \zetabr = \sqrt{\frac{1}{1+\kbr}}, 
$$
where $\betabr$ is the path gain between RIS-BS given by $\betabr = d_{\mathrm{br}}^{-2}/\etabr^2$ and $\myVM{A}{br}{\mathrm{LOS}}$  is given by \eq{\ref{Eq: LOS Component for Hbr}}. The $\myVM{A}{br,c,s}{\mathrm{SC}}$ matrices contain the scattered rays and are calculated in the same manner as $\myVM{A}{d,c,s}{\mathrm{SC}}$ and $\myVM{A}{ru,c,s}{\mathrm{SC}}$. $\kbr$ is the Ricean K-factor for the RIS-BS channel. 
In scenarios where the BS and RIS are located in close proximity, it is reasonable to assume that the RIS-BS channel is dominated by its LOS component \cite{9066923}.

Using \eq{\ref{Eq: Global Channel}} and the channels described above, the received signal at the BS is, 
\begin{equation}\label{Eq: r}
	\mathbf{r} = \mathbf{H}\mathbf{s} + \mathbf{n},
\end{equation}
where $\mathbf{s}$ is a $K \times 1$ vector of transmitted symbols, each with a power of $\Exp{\Abs{s_k}{2}} = E_{s}$ and $\mathbf{n} \sim \mathcal{CN}(0,\sigma^2\mathbf{I}_{M})$. For our results, we will assume without loss of generality that $E_s=1$.

\section{Channel Separation}\label{Sec: Channel Separation}
In this section, we use the channel separation approach \cite{TightBounds} to provide a better understanding of the effect $\myVM{\Phi}{}{}$ has on system performance. Specifically, channel separation separates the global UL channel in \eq{\ref{Eq: Global Channel}} into rows that are explicitly affected by $\myVM{\Phi}{}{}$ and rows that are not. This technique can be used to derive a variety of low complexity RIS designs with very high performance. In this paper, we focus on RIS designs for improving UL sum rate and enhancing the rates achieved by ZF and MMSE receivers. For these three design criteria, the optimization problems can be stated as
\begingroup
\allowdisplaybreaks
\begin{align}
    &R_{\mathrm{sum}}^{\text{opt}} = \maxover{\mathbf{\Phi}} -\log_2\left\lvert ( \sigma^2\mathbf{I}_{K} + {\mathbf{H}}^H{\mathbf{H}} )^{-1} \right\rvert - \log_2(\sigma^{2K}) \label{Eq: SR without CS}, \\
    &R_{\mathrm{ZF}}^{\text{opt}} = \maxover{\mathbf{\Phi}} \sum_{k=1}^{K} \log_2\left( 
	1 + \frac{1}{\sigma^2  \left[\left(\myVM{H}{}{H}\myVM{H}{}{}\right)^{-1} \right]_{kk}} \right) \label{Eq: ZF SR without CS}, \\
	&R_{\mathrm{MMSE}}^{\text{opt}} = 
	\maxover{\mathbf{\Phi}} \sum_{k=1}^{K} \log_2\left( 
	\frac{1}{\sigma^2\left[\left(\sigma^2\mathbf{I}_K + \myVM{H}{}{H}\myVM{H}{}{}\right)^{-1}\right]_{kk}} \right), \label{Eq: MMSE SR without CS}
\end{align}
\endgroup
where the maximization is constrained over the unit amplitude diagonal entries in $\mathbf{\Phi}$. The rate metrics in \eq{\ref{Eq: SR without CS}}-\eq{\ref{Eq: MMSE SR without CS}} are well-known and can be found in \cite{dush,SumRateZF,SumRateMMSE} respectively. The difficulty in finding the optimal RIS phases is largely due to the fact that  $\mathbf{\Phi}$ affects every element of ${\mathbf{H}}$. Hence, the determinant and inverses in \eq{\ref{Eq: SR without CS}}-\eq{\ref{Eq: MMSE SR without CS}} appear to be very complicated functions of $\mathbf{\Phi}$. However, when the RIS-BS link is LOS then $\Hbr{}$ is rank 1 and the RIS phases only affect a rank 1 component of $\mathbf{H}$. Motivated by this observation, we seek to separate this RIS-dependent rank 1 component from the rest of the channel. 
In this section, we assume that the RIS-BS link is pure LOS.

Channel separation is achieved via a unitary transformation of ${\mathbf{H}}$. For any $N \times N$ unitary matrix, ${\mathbf{U}}$, we can define ${\tilde{\mathbf{H}}}={\mathbf{U}}^H{\mathbf{H}}$ and ${\tilde{\mathbf{H}}}^H{\tilde{\mathbf{H}}}={\mathbf{H}}^H{\mathbf{H}}$. Hence, the performance metrics in \eq{\ref{Eq: SR without CS}} - \eq{\ref{Eq: MMSE SR without CS}} are identical when the channel $\mathbf{H}$ is replaced by $\tilde{\mathbf{H}}$. Substituting the expression for $\Hbr{}$ in \eqref{Eq: Hbr pure LOS} into ${\tilde{\mathbf{H}}}$, we obtain
\begin{equation}\label{CS}
{\tilde{\mathbf{H}}}={\mathbf{U}}^H \Hd{} +\sqrt{\beta_{\textrm{br}}}{\mathbf{U}}^H \ab{} \ar{H} \mathbf{\Phi} \Hru{}.
\end{equation}
Note that $\ab{}$ and $\ar{}$ are used in \eqref{CS} as simplified notation for the steering vectors in \eqref{Eq: LOS Component for Hbr} for the $\Hbr{}$ channel. Since $\ar{H} \mathbf{\Phi} \Hru{}$ is a row vector, we can confine the effects of $\mathbf{\Phi}$ to one row of ${\tilde{\mathbf{H}}}$ by selecting ${\mathbf{U}}$ to satisfy 
\begin{equation}\label{CS_sol}
{\mathbf{U}}^H \ab{} = \sqrt{M}[1, 0, \ldots ,0]^T.
\end{equation}
The unitary matrix satisfying \eqref{CS_sol} is the matrix of left singular vectors of $\Hbr{}$ as shown below.

Define the singular value decomposition (SVD) of $\Hbr{}$ as 
$\Hbr{} = \myVM{U}{}{}\myVM{D}{}{}\myVM{V}{}{H}$, 
where $\myVM{U}{}{}=[\myVM{u}{1}{}, \ldots \mathbf{u}_{M}]$ is the matrix of left singular vectors, $\myVM{D}{}{}$ is the diagonal matrix of singular values and $\myVM{V}{}{}=[\myVM{v}{1}{}, \ldots \mathbf{v}_{N}]$  is the matrix of right singular vectors. Since $\Hbr{}$ is rank-1, then only one non-zero singular value, $d_1$, exists and $\Hbr{} =d_1\myVM{u}{1}{}\myVM{v}{1}{H}$ where $\myVM{u}{1}{} = \ab{}/\sqrt{M}$, $\myVM{v}{1}{} = \ar{}/\sqrt{N}$ and $d_1 = \sqrt{M N \betabr}$. 
Using this value of $\myVM{U}{}{}$, we have
\begin{align}
\tilde{\mathbf{H}} &=  \Ubr{H}\Hd{} + 
	\begin{bmatrix}
		\ab{H}/\sqrt{M} \notag \\
		\myVM{u}{2}{H} \notag \\
		\vdots \notag \\
		\myVM{u}{M}{H}
	\end{bmatrix}
	\sqrt{\betabr}\ab{}\ar{H}\myVM{\Phi}{}{}\Hru{} \notag \\
	&= 
	\begin{bmatrix}
		\myVM{u}{1}{H}\Hd{} + \sqrt{M\betabr}\ar{H}\myVM{\Phi}{}{}\Hru{} \notag \\
		\myVM{u}{2}{H}\Hd{} \notag \notag \\
		\vdots \notag \\
		\myVM{u}{M}{H}\Hd{}
	\end{bmatrix} \notag \\
	&\triangleq 
	\begin{bmatrix}
		\wwr{H} \\
		\myVM{H}{1}{}
	\end{bmatrix}.\label{chan_sep}
\end{align}
Channel separation is observed in \eqref{chan_sep} as $\wwr{H}$, the first row of ${\tilde{\mathbf{H}}}$, is the only row affected by $\myVM{\Phi}{}{}$. Hence, we can rewrite the sum rate metrics in \eq{\ref{Eq: SR without CS}} - \eq{\ref{Eq: MMSE SR without CS}} in terms of the vector $\mathbf{w}$ and $\mathbf{H}_1$ given in \eqref{chan_sep}.

Next, we derive alternative expressions for \eq{\ref{Eq: SR without CS}} - \eq{\ref{Eq: MMSE SR without CS}} which only depend on the RIS phases through the $\wwr{}$ vector and allow closed form RIS designs to be derived. Noting that the common term in \eq{\ref{Eq: SR without CS}} - \eq{\ref{Eq: MMSE SR without CS}} is of the form $(\alpha \mathbf{I}_K + \myVM{H}{}{H}\myVM{H}{}{})^{-1}$ where $\alpha \in \{0,\sigma^2 \}$, we rewrite this term using the matrix inversion lemma to give
\begin{align}\label{Eq: S(X)}
	&\left(\alpha \mathbf{I}_K + \myVM{H}{}{H}\myVM{H}{}{}\right)^{-1} \notag \\
	& = 
	\left(\alpha \mathbf{I}_K + \myVM{\tilde{H}}{}{H}\myVM{\tilde{H}}{}{}\right)^{-1} \notag \\
	&=
	\left(\alpha \mathbf{I}_K + \myVM{H}{1}{H}\myVM{H}{1}{} + \myVM{w}{}{}\myVM{w}{}{H} \right)^{-1} \notag \\
	&\triangleq
	\left(\mathbf{Q} + \myVM{w}{}{}\myVM{w}{}{H} \right)^{-1} \notag \\
	&=
	\myVM{Q}{}{-1} - \myVM{Q}{}{-1}\myVM{w}{}{}\left(1 + \myVM{w}{}{H}\myVM{Q}{}{-1}\myVM{w}{}{}\right)^{-1}\myVM{w}{}{H}\myVM{Q}{}{-1} \notag \\
	& \triangleq \mathbf{S(Q)},
\end{align}
where $\mathbf{Q}$ is a Hermitian matrix and its formulation is specific to the metric being optimized.
In deriving \eqref{Eq: S(X)}, the SVD of the $M \times N$ matrix $\Hbr{}$ was used. However, the final solution can be written in terms of the channels only, making it computationally trivial involving only a $K \times K$ determinant and a $K \times K$ inverse. This is achieved by writing $\myVM{U}{}{}=[\myVM{u}{1}{} \myVM{U}{2}{}]$, so that  $\myVM{U}{}{} \myVM{U}{}{H} =  \myVM{I}{M}{} =\myVM{u}{1}{} \myVM{u}{1}{H}+\myVM{U}{2}{} \myVM{U}{2}{H}$. Using this result and substituting $\myVM{H}{1}{} = \myVM{U}{2}{H}\Hd{}$ and $\myVM{u}{1}{} = \ab{}/\sqrt{M}$ gives
\begingroup
\allowdisplaybreaks
\begin{align*}
    \mathbf{Q} &= \alpha \mathbf{I}_K + \myVM{H}{1}{H}\myVM{H}{1}{} \\
    &= \alpha \mathbf{I}_K + 
    \myVM{H}{d}{H}\myVM{U}{2}{}\myVM{U}{2}{H}\myVM{H}{d}{} \\
    &= \alpha\mathbf{I}_K+\myVM{H}{d}{H}(\myVM{I}{M}{}-\myVM{u}{1}{}\myVM{u}{1}{H})\myVM{H}{d}{} \\
    &= \alpha\mathbf{I}_K+\myVM{H}{d}{H}(\myVM{I}{M}{}-\ab{}\ab{H}/M)\myVM{H}{d}{}.
\end{align*}
\endgroup
Hence, the $\mathbf{Q}$ matrices for the three optimization problems are
\begin{align}
    \eq{\ref{Eq: SR without CS}} :\, &\myVM{Q}{Sum}{}
    =\sigma^2\mathbf{I}_K+\myVM{H}{d}{H}(\myVM{I}{M}{}-\ab{}\ab{H}/M)\myVM{H}{d}{}\label{Eq: Q Matrix for SR}, \\
    \eq{\ref{Eq: ZF SR without CS}} :\, &\myVM{Q}{ZF}{} 
    =\myVM{H}{d}{H}(\myVM{I}{M}{}-\ab{}\ab{H}/M)\myVM{H}{d}{}\label{Eq: Q Matrix for ZF}, \\
    \eq{\ref{Eq: MMSE SR without CS}} :\, &\myVM{Q}{MMSE}{} = \myVM{Q}{Sum}{} \label{Eq: Q Matrix for MMSE}.
\end{align}
Using \eqref{Eq: S(X)}, the optimization problems can be equivalently written as
\begingroup
\allowdisplaybreaks
\begin{align}
    R_{\mathrm{Sum}}^{\text{opt}} &= \maxover{\mathbf{\Phi}} -\log_2\left\lvert \mathbf{S(\myVM{Q}{Sum}{})} \right\rvert - \log_2(\sigma^{2K}) \label{Eq: SR with CS}, \\
    R_{\mathrm{ZF}}^{\text{opt}} &= \maxover{\mathbf{\Phi}} \sum_{k=1}^{K} \log_2\left( 
	1 + \frac{1}{\sigma^2 \left[ \mathbf{S(\myVM{Q}{ZF}{})} \right]_{kk}} \right) \label{Eq: ZF SR with CS}, \\
	R_{\mathrm{MMSE}}^{\text{opt}} &= 
	\maxover{\mathbf{\Phi}} \sum_{k=1}^{K} \log_2\left( 
	\frac{1}{\sigma^2 \left[\mathbf{S(\myVM{Q}{MMSE}{})}\right]_{kk}} \right) \label{Eq: MMSE SR with CS}.
\end{align}
\endgroup
The matrices, $\mathbf{S(\cdot)}$, in \eqref{Eq: SR with CS} - \eqref{Eq: MMSE SR with CS} are functions of the RIS phases only through the vector, $\myVM{w}{}{}$, defined by
\begin{equation}\label{Eq: v vector}
    \myVM{w}{}{} =\Hd{H}\myVM{a}{b}{}/\sqrt{M} + \sqrt{M\betabr}\Hru{H}\myVM{\Phi}{}{H}\ar{}. 
\end{equation}
This is an important result of channel separation as the RIS design has now collapsed to optimizing the vector, $\myVM{w}{}{}$.

In the next section, we develop low-complexity RIS designs for these optimization problems. The methods are separated into the two important scenarios where the RIS phases are discrete (Sec.~\ref{SubSec: Low bit RIS phases}) and when the RIS phases are continuous (Sec.~\ref{SubSec: High bit RIS phases}).

\section{Optimization: Discrete and Continuous Phases}\label{Sec: Optimization} 
Here, we propose low-complexity approaches to the maximizations of $R_{\mathrm{Sum}}, R_{\mathrm{ZF}}, R_{\mathrm{MMSE}}$ given in Sec.~\ref{Sec: Channel Separation} for two different scenarios; the RIS phases are either discrete or continuous. Note that the term 'discrete' refers to the quantization of the RIS phases where we use the terminology 'low-bit phase resolution' to signify low level quantization and 'high-bit phase resolution' for high level quantization. Here, the number of bits used to quantize the RIS phases is denoted by $b$.

\subsection{Discrete RIS Phases}\label{SubSec: Low bit RIS phases}
In many implementations of RIS-aided wireless systems, it is appropriate to assume that the phase of each RIS element is selected from a finite number of phases (i.e. discrete RIS phases). Here, we propose an alternating optimization algorithm to maximize $R_{\mathrm{Sum}}, R_{\mathrm{ZF}}, R_{\mathrm{MMSE}}$ for discrete RIS phases. Note that the AO algorithm is not intended to be a numerical procedure to fully  optimize performance. Rather, it is used as a vehicle to to achieve a low complexity solution, suitable for practical implementation as the number of iterations is heavily constrained.

Since the effect of the RIS phases has been reduced to a single vector (see $\mathbf{w}$ in Sec.~\ref{Sec: Channel Separation}), it is now feasible to design the RIS phases by iterating through the $N$ RIS elements and searching over the set of possible discrete elements. Firstly, since the steering vector, $\ar{}$, satisfies the unit amplitude constraint, we can write the RIS matrix as $\mathbf{\Phi} = \Diag{\ar{}}\mathbf{\Gamma}$, where $\mathbf{\Gamma}$ is a modified diagonal phase matrix. Note that the optimization can now proceed over the $\mathbf{\Gamma}$ matrix or over $\myVM{x}{D}{H} = [e^{j\gamma_1},\ldots,e^{j\gamma_N}]$ where $\myVM{x}{D}{}$ is a $N \times 1$ vector containing the diagonal elements of $\mathbf{\Gamma}$.  The elements of $\myVM{x}{D}{}$ are chosen by selecting $\gamma_i$, $i=1,2, \ldots N$ from the discrete set,
\begin{equation}\label{Eq: Discrete set of phi_n}
	\mathcal{S} = \left\{ 0, \frac{2\pi}{2^b}, \frac{4\pi}{2^b}, \ldots, \frac{2\pi(2^b-1)}{2^b} \right\}.
\end{equation}
Hence, the RIS phases are selected from $\mathcal{S}$ with a phase offset given by the $\myVM{a}{r}{}$ vector.
With this notation, the conjugate transpose of the vector $\mathbf{w}$ can be written as,
\begin{equation}
	 \mathbf{w}^{H} = \frac{\ab{H}}{\sqrt{M}}\Hd{} + \sqrt{M\betabr}\myVM{x}{D}{H}\Hru{}. \label{Eq: f(x)}
\end{equation}
Utilising this result, we can iteratively optimize the system for any of the sum rate metrics in Sec.~\ref{Sec: Channel Separation}. We first compute an initial starting point for the algorithm, 
which is to compute the sum rate metric from the phase vector $\myVM{x}{D}{(0)} = [1,\ldots,1]^T$. Note that other initial points could be used but we use the simplest possible.  
We then iterate through each RIS element, finding the phase from the set \eqref{Eq: Discrete set of phi_n} which causes the largest increase in the sum rate metric. 
As an example, we provide the layout of the algorithm for maximizing $R_{\mathrm{ZF}}$ in Algorithm~\ref{Algorithm: MUIQ}.
\begin{algorithm}[h]
	\SetAlgoLined
	Set $\mathbf{Q}^{\mathrm{ZF}} = \Hd{H}\left(\mathbf{I}_M - \frac{\ab{}\ab{H}}{M} \right)\Hd{} $. \\
	Set $\myVM{x}{D}{(0)} = [1,\ldots,1]^T$. \\
	Set $\myVM{a}{}{H} = \frac{\myVM{a}{b}{H}\myVM{H}{d}{}}{\sqrt{M}}$ and $\mathbf{B} = \sqrt{M\betabr}\myVM{H}{ru}{}$ \\
	Compute $\mathbf{w}^{(0)} = \ \myVM{a}{}{} + \myVM{B}{}{H}(\myVM{x}{D}{(0)}) $. \\
	Compute $\mathbf{S}\left( \myVM{Q}{ZF}{} \right)$ using $\mathbf{w}^{(0)}$ and $\myVM{Q}{ZF}{}$\\
	Compute $R_{ZF}^{(0)}$\\
	Set $k=1$. \\
	Set $l=1$ and set $L$ to be the number of iterations. \\
	\While{$l \leq L$}{
	\For{$n=1 : N$}{
	    Set $\myVM{y}{}{} = \myVM{x}{D}{(k-1)}$ \\
		\For{$l=1 : 2^b$}{
			Set $\gamma_{n}$ to be the $l^{\mathrm{th}}$ element from the set \eq{\ref{Eq: Discrete set of phi_n}}.\\
			Set the $n^{\mathrm{th}}$ element in $y_n = e^{j\gamma_{n}}$.\\
			Compute $\mathbf{w}^{(k)} =  \myVM{a}{}{} + \myVM{B}{}{H}\myVM{y}{}{} $. \\
			Compute $\mathbf{S}\left( \myVM{Q}{ZF}{} \right)$ using $\mathbf{w}^{(k)}$ and $\myVM{Q}{ZF}{}$\\
	        Compute $R_{ZF}^{(k)}$\\
			\If{$R_{ZF}^{(k)} \geq R_{ZF}^{(k-1)}$}{
				Set $x_{\mathrm{D},n} = e^{j\gamma_{n}}$\\
			}
			Set $k = k + 1$.
		}
	}
	$l=l+1$
	}
	Return $\myVM{x}{D}{}$.
	\caption{MUIQ: Multi-User Iterative Quantisation}
	\label{Algorithm: MUIQ}
\end{algorithm}

In Algorithm~\ref{Algorithm: MUIQ}, $L$ is the number of repeats of the procedure. Note that Algorithm~\ref{Algorithm: MUIQ} can be used to optimize any of the given metrics in Sec.~\ref{Sec: Channel Separation}, with the difference being in the $\mathbf{Q}$ matrix, which is selected for the metric being optimized in Sec.~\ref{Sec: Channel Separation}. It is worth noting that for minimizing a metric, the inequality in the decision step of the algorithm is inverted. 


The computational complexity of Algorithm~\ref{Algorithm: MUIQ} is dominated by the computation of $\myVM{B}{}{H}\myVM{y}{}{}$ in $\myVM{w}{}{(k)}$ which grows as $\mathcal{O}\left( KN \right)$. Hence, due to the repeated computation over $N$ RIS elements and $2^b$ possible RIS phases, the overall computational complexity of Algorithm~\ref{Algorithm: MUIQ} is $\mathcal{O}\left( L 2^b K N^2 \right)$. 

Algorithm~\ref{Algorithm: MUIQ} is therefore a useful approach to finding RIS designs to maximize $R_{\mathrm{Sum}}, R_{\mathrm{ZF}}, R_{\mathrm{MMSE}}$ for scenarios where the quantization level $b$ is low and when the procedure is not frequently repeated (i.e small $L$). However, when the quantization level of the RIS phases is high or in scenarios where the RIS phases are continuous, a different approach is required, which is covered in the next section.




\subsection{Continuous RIS Phases}\label{SubSec: High bit RIS phases}

In this section, we consider the case where the RIS phases can be chosen from any continuous value in $[0,2\pi]$. First, we present a simple closed form solution to approximate the maximization of $R_{\mathrm{sum}}$ (Sec.~\ref{SubSubSec: Max RSum}). Next, we consider an approach to enhance the performance of MMSE and ZF receivers..


\subsubsection{$R_{\mathrm{Sum}}$}\label{SubSubSec: Max RSum}
For ease of notation, we let  $\mathbf{P}=(\mathbf{Q}_{\mathrm{Sum}})^{-1}$. Substituting the formula for $\mathbf{S}(\mathbf{P})$ given by \eqref{Eq: S(X)} into the sum rate expression \eq{\ref{Eq: SR with CS}}, we have 
\begingroup
\allowdisplaybreaks
\begin{align}\label{Eq: Sum rate expanded form}
	& R_{\mathrm{Sum}} \notag \\
	&= -\log_2\left( \lvert \mathbf{P} - \mathbf{P}\myVM{w}{}{}(1+\myVM{w}{}{H}\mathbf{P}\myVM{w}{}{})^{-1}\myVM{w}{}{H}\mathbf{P} \rvert \right)
	- \log_2(\sigma^{2K}) \notag \\
	&\overset{(a)}{=} -\log_2\left( \Det{\mathbf{P}}(1 - (1 + \myVM{w}{}{H}\myVM{P}{}{}\myVM{w}{}{})^{-1}\myVM{w}{}{H}\myVM{P}{}{}\myVM{w}{}{}) \right) 
	- \log_2(\sigma^{2K}) \notag \\ 
	&= -\log_2\left( \Det{\mathbf{P}} \right) - \log_2\left( 1 - \frac{\myVM{w}{}{H}\myVM{P}{}{}\myVM{w}{}{}}{1 + \myVM{w}{}{H}\myVM{P}{}{}\myVM{w}{}{}} \right) 
	- \log_2(\sigma^{2K}) \notag \\
	&= -\log_2\left( \Det{\mathbf{P}} \right) + 
	\log_2\left( 1 + \myVM{w}{}{H}\mathbf{P}\myVM{w}{}{}  \right) - \log_2(\sigma^{2K}),
\end{align}
\endgroup
where in $(a)$ we utilize the Matrix Determinant Lemma. Finding the maximum of \eq{\ref{Eq: Sum rate expanded form}} is equivalent to maximizing $\myVM{w}{}{H}\mathbf{P}\myVM{w}{}{}$ where $\mathbf{w}$ is given in \eqref{Eq: v vector}. Let $\mathbf{x}^H = [e^{j\phi_1},\ldots,e^{j\phi_N}]$ be the vector of RIS phases, $\myVM{w}{1}{} = \Hd{H}\ab{}/\sqrt{M}$, $\mathbf{A}_1 = \sqrt{M\betabr}\Diag{\ar{H}}\Hru{}$, then
\begin{align}
    \myVM{w}{}{H}\myVM{P}{}{}\myVM{w}{}{}
    =
    \myVM{w}{1}{H}\myVM{P}{}{}\myVM{w}{1}{}
    +
    \myVM{x}{}{H}\myVM{A}{1}{}\myVM{P}{}{}\myVM{A}{1}{H}\myVM{x}{}{}
    +
    2\Re\{ \myVM{x}{}{H}\myVM{A}{1}{}\myVM{P}{}{}\myVM{w}{1}{} \} . \label{Eq: wH P w}
\end{align}
Note that the first two terms in \eqref{Eq: wH P w} are quadratic and dominate the third term. This is further accentuated by any maximizing of the terms over the RIS phases. To motivate the dominance of the quadratic terms further, consider the third term in \eqref{Eq: wH P w} which can be written as $2\Re\{ \myVM{x}{}{H}\myVM{y}{}{} \}$ where $\myVM{y}{}{} = \sqrt{\betabr}\Diag{\myVM{a}{r}{H}}\Hru{}\myVM{P}{}{}\Hd{H}\myVM{a}{b}{}$. Even if the RIS design only optimises this term, the maximum value obtained is $2\sum_{n=1}^{N} \Abs{y_n}{}$ which is $\mathcal{O}(N)$. In contrast, the second quadratic term given by $\mathbf{x}^H\myVM{A}{1}{}\myVM{P}{}{}\myVM{A}{1}{H}\mathbf{x}$ can approach $N \lambda_{\mathrm{max}}(\myVM{A}{1}{}\myVM{P}{}{}\myVM{A}{1}{H})$ if $\myVM{x}{}{}$ is chosen to match the phases of the maximum eigenvector of $\myVM{A}{1}{}\myVM{P}{}{}\myVM{A}{1}{H}$. Using the definition of $\myVM{A}{1}{}$, we obtain $N \lambda_{\mathrm{max}}(\myVM{A}{1}{}\myVM{P}{}{}\myVM{A}{1}{H}) = NM\betabr \lambda_{\mathrm{max}}(\Hru{}\myVM{P}{}{}\Hru{H})$. Typically, the maximum eigenvalue of $\Hru{}\myVM{Q}{}{-1}\Hru{H}$ is $\mathcal{O}(N)$ so the quadratic term grows as $\mathcal{O}(N^2)$. As a result, the quadratic terms combined are of the order of $N$ times larger than the cross product term. Hence, as an approximation, we have
\begin{align*}
	\myVM{w}{}{H}\mathbf{P}\myVM{w}{}{} \approx \mathbf{x}^{H}\left( \mathbf{A}_1\mathbf{P}\mathbf{A}_1^H + \nu\mathbf{I}_N \right)\mathbf{x}
	\triangleq \mathbf{x}^{H}\mathbf{Z}\mathbf{x},
\end{align*}
where $\nu = \frac{\myVM{w}{1}{H}\mathbf{P}\myVM{w}{1}{}}{N}$. The optimization problem can therefore be formulated as
\begin{equation}\label{Eq: P.1}
    \begin{aligned}
          \argmax{\mathbf{x}}  \quad & \mathbf{x}^H \mathbf{Z} \mathbf{x} \\
        \text{s.t.} \quad &\Abs{x_{i}}{}=1 \text{ for } i=1,\ldots,N.
    \end{aligned}
    \tag{P.3}
\end{equation}
Notice that if the unit amplitude constraint on the RIS phases is relaxed to $\mathbf{x}^H\mathbf{x}=N$, then the optimum solution, $\mathbf{x}^{\star}$, is proportional to the maximal eigenvector of $\mathbf{Z}$. Direct computation of $\mathbf{x}^{\star}$ requires the eigenvalue decomposition of an $N \times N$ matrix. Alternatively, we use App.~\ref{App: finding max eigenvector} as a low complexity approach to computing $\mathbf{x}^{\star}$, which gives
\begin{equation}\label{Eq: maxeig}
	\mathbf{x}^{\star} = \mathbf{A}_1 \mathbf{x'}^{\star},
\end{equation}
where $\mathbf{x'}^{\star} \propto \text{max eigenvector}\left\{ \nu\mathbf{I}_K + \mathbf{P}\mathbf{A}_1^H\mathbf{A}_1 \right\}$. The problem has been reduced from an $N \times N$ to a $K \times K$ eigenvalue decomposition, a considerable saving especially when considering large RIS sizes. However, this approach does not restrict $x_{i}^{\star}$ to unit amplitude (i.e. $\Abs{x_{i}^{\star}}{} = 1$). To resolve this issue, we consider the alternative problem of finding the unit amplitude vector as close as possible to the maximum eigenvector. Specifically, we minimize the $\ell_1$-norm of the residuals between $\mathbf{x}^{\star}$ in \eqref{Eq: maxeig} and the solution to the relaxed version of \eqref{Eq: P.1}. Mathematically, the alternative optimization problem is
\begin{equation}\label{Eq: P.2}
    \begin{aligned}
        \min \quad & \Norm{\mathbf{x}^{\star} - \hat{\mathbf{x}}}{1}{} = \Abs{x^{\star}_1 - \hat{x}_1}{} + \ldots + \Abs{x^{\star}_N - \hat{x}_N}{}\\
        \text{s.t.} \quad &\Abs{\hat{x}_i}{}=1 \text{ for } i=1,\ldots,N.
    \end{aligned}
    \tag{P.4}
\end{equation}
The solution to \eq{\ref{Eq: P.2}} is given in \cite{TightBounds} where 
\begin{equation}\label{Eq: Closed form optimal RIS}
    \hat{\mathbf{x}} = [e^{j \angle x^{\star}_1}, \ldots, e^{j \angle x^{\star}_N}]^T.
\end{equation}
Thus, using the phases in \eqref{Eq: Closed form optimal RIS} is a well-motivated approximation to the maximization of $R_{\mathrm{Sum}}$ in scenarios where the RIS phases are continuous.
The computation of \eqref{Eq: Closed form optimal RIS} is dominated by the eigenvalue decomposition of a $K \times K$ matrix and the inverse of the $K \times K$ matrix $\mathbf{Q}$. Hence, the computational complexity of \eqref{Eq: Closed form optimal RIS} is $\mathcal{O}(K^3)$. 

In summary, we can use \eqref{Eq: Closed form optimal RIS} as an approximate solution to the sum rate maximization problem \eqref{Eq: SR with CS}. In this paper, we refer to \eqref{Eq: Closed form optimal RIS} as the sum rate solution.

\subsubsection{Total Mean Square Error}\label{SubSubSec: Min Total MSE}
In this section, we consider the design of continuous RIS phases to enhance the performance of MMSE and ZF receivers. A direct attempt to maximize the sum rates in \eqref{Eq: ZF SR with CS} and \eqref{Eq: MMSE SR with CS} appears very challenging due to the summation of logarithmic terms. Hence, we target a related but simpler metric, the total mean squared error, $\mathrm{MSE}_{\mathrm{Tot}}$, of the MMSE receiver. As ZF and MMSE receivers behave similarly at high SNR, we also use this design for ZF receivers. The total  mean squared error to be minimized is defined by \cite{668742} 
\begin{equation}\label{Eq: MSE metric}
    \mathrm{MSE}_{\mathrm{Tot}} = \sum_{k=1}^{K}
    \Exp{\Abs{s_{k} - \hat{s}_{k}}{2}}
    = 
    \tr{\mathbf{S}(\mathbf{Q}_{\mathrm{MMSE}})}{}{},
\end{equation}
where $\hat{s}_k$ is the $k^{\mathrm{th}}$ estimated transmitted symbol. Note that just as with the sum rate metrics in \eqref{Eq: SR with CS}-\eqref{Eq: MMSE SR with CS}, we can write the total MSE in terms of $\mathbf{S}\left( \cdot \right)$. This is the key observation as writing  $\mathrm{MSE}_{\mathrm{Tot}}$ in this form allows channel separation to be applied to the problem.

Firstly, we expand the total MSE expression and using the properties of the $\tr{\cdot}{}{}$ operator, 
\begin{align}\label{Eq: Total MSE expanded form}
	\mathrm{MSE}_{\mathrm{Tot}} &= \mathrm{tr}\left\{\mathbf{P} - \mathbf{P}\myVM{w}{}{}(1 + \myVM{w}{}{H}\mathbf{P}\myVM{w}{}{})^{-1}\myVM{w}{}{H}\mathbf{P} \right\} \notag \\
	&= \tr{\mathbf{P}}{}{} - \frac{\myVM{w}{}{H}\mathbf{P}^2\myVM{w}{}{}}{1 + \myVM{w}{}{H}\mathbf{P}\myVM{w}{}{}} \notag \\
	&\triangleq \tr{\mathbf{P}}{}{} - T.
\end{align}
where for ease of notation, we let  $\mathbf{P}=(\mathbf{Q}_{\mathrm{MMSE}})^{-1}$.
Hence, minimizing the total MSE is equivalent to maximizing $T$ in \eq{\ref{Eq: Total MSE expanded form}}. 
As in \eqref{Eq: wH P w}, we use $\mathbf{x}^H = [e^{j\phi_1},\ldots,e^{j\phi_N}]$ and substitute $\mathbf{w}$ from \eqref{Eq: v vector} into the numerator and denominator of $T$ to give
\begin{align}
    \myVM{w}{}{H}\myVM{P}{}{2}\myVM{w}{}{}
    &=
    \myVM{w}{1}{H} \myVM{P}{}{2} \myVM{w}{1}{}
    +
    \myVM{x}{}{H}\myVM{A}{1}{}\myVM{P}{}{2}\myVM{A}{1}{H}\myVM{x}{}{}
    +
    2\Re\{ \myVM{x}{}{H}\myVM{A}{1}{}\myVM{P}{}{2}\myVM{w}{1}{} \}. \label{Eq: wH P2 w}
\end{align}
\begin{align}
    1 + \myVM{w}{}{H}\myVM{P}{}{}\myVM{w}{}{} &=
    ( 1 + \myVM{w}{1}{H} \myVM{P}{}{} \myVM{w}{1}{} )
    +
    \myVM{x}{}{H}\myVM{A}{1}{}\myVM{P}{}{}\myVM{A}{1}{H}\myVM{x}{}{} \notag \\
    & \quad +
    2\Re\{ \myVM{x}{}{H}\myVM{A}{1}{}\myVM{P}{}{}\myVM{w}{1}{} \}. \label{Eq: 1 + wH P w}
\end{align}
Just as in \eqref{Eq: wH P w}, where we motivate the approximation of $\myVM{w}{}{H}\myVM{P}{}{}\myVM{w}{}{}$ by only including the dominating quadratic terms, we also approximate \eqref{Eq: wH P2 w} and \eqref{Eq: 1 + wH P w} as follows,
\begin{align}
    \myVM{w}{}{H}\myVM{P}{}{2}\myVM{w}{}{}
    &\approx
    \myVM{w}{1}{H} \myVM{P}{}{2} \myVM{w}{1}{}
    +
    \myVM{x}{}{H}\myVM{A}{1}{}\myVM{P}{}{2}\myVM{A}{1}{H}\myVM{x}{}{}
    \notag \\
    & = \myVM{x}{}{H}\left( 
    \myVM{A}{1}{}\myVM{P}{}{2}\myVM{A}{1}{H}
    +
    \frac{\myVM{w}{1}{H} \myVM{P}{}{2} \myVM{w}{1}{}}{N} \mathbf{I}_N
    \right)\myVM{x}{}{}. \label{Eq: wH P2 w approx}
\end{align}
\begin{align}
    1 + \myVM{w}{}{H}\myVM{P}{}{}\myVM{w}{}{} 
    &\approx
    ( 1 + \myVM{w}{1}{H} \myVM{P}{}{} \myVM{w}{1}{} )
    +
    \myVM{x}{}{H}\myVM{A}{1}{}\myVM{P}{}{}\myVM{A}{1}{H}\myVM{x}{}{} \notag \\
    &=
    \myVM{x}{}{H} \left(
    \myVM{A}{1}{}\myVM{P}{}{}\myVM{A}{1}{H}
    +
    \frac{1 + \myVM{w}{1}{H} \myVM{P}{}{} \myVM{w}{1}{}}{N} \mathbf{I}_N
    \right)
    \myVM{x}{}{}. \label{Eq: 1 + wH P w approx}
\end{align}
Using \eqref{Eq: wH P2 w approx} and \eqref{Eq: 1 + wH P w approx}, we have
\begingroup
\allowdisplaybreaks
\begin{align*}
    T
    \approx
    \frac{\myVM{x}{}{H}\left( 
    \myVM{A}{1}{}\myVM{P}{}{2}\myVM{A}{1}{H}
    +
    \frac{\myVM{w}{1}{H} \myVM{P}{}{2} \myVM{w}{1}{}}{N} \mathbf{I}_N
    \right)\myVM{x}{}{}}{\myVM{x}{}{H} \left(
    \myVM{A}{1}{}\myVM{P}{}{}\myVM{A}{1}{H}
    +
    \frac{1 + \myVM{w}{1}{H} \myVM{P}{}{} \myVM{w}{1}{}}{N} \mathbf{I}_N
    \right)
    \myVM{x}{}{}}
    \triangleq
    \frac{\alpha_1 \mathbf{x}^H\mathbf{Z}_1\mathbf{x}}{\alpha_2 \mathbf{x}^H\mathbf{Z}_2\mathbf{x}}, 
\end{align*}
\endgroup
with
\begin{equation*}
	\mathbf{Z}_1 = \frac{1}{\alpha_1}\myVM{A}{1}{}\mathbf{P}^2\myVM{A}{1}H + \mathbf{I}_N, \quad
	\mathbf{Z}_2 = \frac{1}{\alpha_2}\myVM{A}{1}{}\mathbf{P}\myVM{A}{1}{H} + 
	\mathbf{I}_N,
\end{equation*}
where $\alpha_1 = \frac{\myVM{w}{1}{H}\myVM{P}{}{2}\myVM{w}{1}{}}{N}$ and $\alpha_2 = \frac{(1+\myVM{w}{1}{H}\myVM{P}{}{}\myVM{w}{1}{})}{N}$. As $\alpha_1 > 0, \alpha_2 > 0$ are independent of $\mathbf{x}$, we 
approximate the minimization of $\mathrm{MSE}_{\mathrm{Tot}}$ by the following optimization problem
\begin{equation}\label{Eq: P.5}
    \begin{aligned}
          \argmax{\mathbf{x}}  \quad & \frac{\mathbf{x}^H\mathbf{Z}_1\mathbf{x}}{\mathbf{x}^H\mathbf{Z}_2\mathbf{x}} \\
        \text{s.t.} \quad &\Abs{x_{i}}{}=1 \text{ for } i=1,\ldots,N.
    \end{aligned}
    \tag{P.5}
\end{equation}

From \cite{6200372}, the solution to \eq{\ref{Eq: P.5}} can be found using an eigenvalue decomposition. Specifically, we have $\mathbf{x}^{\star} \propto \text{max eigenvector}\left\{\mathbf{Z}_2^{-1}\mathbf{Z}_1\right\}$ as the solution. Notice that this would require an $N \times N$ inverse and an eigenvalue decomposition of a $N \times N$ matrix, which is very expensive for large RIS sizes. This can be drastically reduced to the inverse and eigenvalue decomposition of a $K \times K$ matrix as follows. Using the matrix inverse lemma, we have
\begin{equation*}
	\mathbf{Z}_2^{-1} = \mathbf{I}_N - \myVM{A}{1}{}\left(\alpha_2\mathbf{P}^{-1} + \myVM{A}{1}{H}\myVM{A}{1}{}\right)^{-1}\myVM{A}{1}{H}.
\end{equation*}
Then $\mathbf{Z}_2^{-1}\mathbf{Z}_1$ results in
\begingroup
\allowdisplaybreaks
\begin{align*}
	\mathbf{Z}_2^{-1}\mathbf{Z}_1
	&= \left(  \mathbf{I}_N - \myVM{A}{1}{}\left(\alpha_2\mathbf{P}^{-1} + \myVM{A}{1}{H}\myVM{A}{1}{}\right)^{-1}\myVM{A}{1}{H} \right) \\
	& \quad \times \left( \mathbf{I}_N + \frac{1}{\alpha_1}\myVM{A}{1}{}\mathbf{P}^2\myVM{A}{1}{H} \right) \notag \\
	&= \mathbf{I}_N + \frac{\myVM{A}{1}{}\myVM{P}{}{2}\myVM{A}{1}{H}}{\alpha_1}
	- \myVM{A}{1}{}\left( \alpha_2\myVM{P}{}{-1} + \myVM{A}{1}{H}\myVM{A}{1}{} \right)^{-1}\myVM{A}{1}{H} \notag \\
	& \quad - \frac{1}{\alpha_1} \myVM{A}{1}{}\left( \alpha_2\myVM{P}{}{-1} + \myVM{A}{1}{H}\myVM{A}{1}{} \right)^{-1} \myVM{A}{1}{H}\myVM{A}{1}{}\myVM{P}{}{2}\myVM{A}{1}{H} \notag \\
	&= \mathbf{I}_N + \myVM{A}{1}{}\Bigg(
	\frac{\myVM{P}{}{2}}{\alpha_1}
	-
	\left( \alpha_2\myVM{P}{}{-1} + \myVM{A}{1}{H}\myVM{A}{1}{} \right)^{-1}
	\notag \\
	& \quad - \frac{1}{\alpha_1}
	\left( \alpha_2\myVM{P}{}{-1} + \myVM{A}{1}{H}\myVM{A}{1}{} \right)^{-1}
	\myVM{A}{1}{H}\myVM{A}{1}{}\myVM{P}{}{2}
	\Bigg)\myVM{A}{1}{H} \notag \\
	&= \mathbf{I}_N + \myVM{A}{1}{}\myVM{Z}{3}{}\myVM{A}{1}{H},
\end{align*}
\endgroup
where, after some algebraic simplification, $\myVM{Z}{3}{} = \left(\alpha_2\mathbf{P}^{-1} + \myVM{A}{1}{H}\myVM{A}{1}{}\right)^{-1}
\left( \frac{\alpha_2}{\alpha_1}\mathbf{P} - \mathbf{I}_K \right)$. A low complexity approach to computing the maximum eigenvector of $\mathbf{I}_N + \myVM{A}{1}{}\myVM{Z}{3}{}\myVM{A}{1}{H}$ is given in App.~\ref{App: finding max eigenvector}, which gives
\begin{equation}\label{Eq: Optimum w}
	\mathbf{x}^{\star} = \myVM{A}{1}{} \mathbf{x'}^{\star},
\end{equation}
where $\mathbf{x'}^{\star} \propto \text{max eigenvector}\left\{ \mathbf{I}_K + \myVM{Z}{3}{}\myVM{A}{1}{H}\myVM{A}{1}{} \right\}$. However, as in \eqref{Eq: maxeig}, this solution does not have the unit amplitude constraint. To resolve this problem, we adopt the approach in \eqref{Eq: P.2}. Specifically, we minimize the $\ell_1$-norm of the residuals between $\mathbf{x}^{\star}$ in \eqref{Eq: Optimum w} and the solution to the relaxed version of \eqref{Eq: P.5}. Mathematically, the alternative optimization problem is given by \eqref{Eq: P.2} where $\myVM{x}{}{\star}$ is given by \eqref{Eq: Optimum w}, which gives the solution as,
\begin{equation}\label{Eq: Closed form optimal RIS total MSE}
    \hat{\mathbf{x}} = [e^{j \angle x^{\star}_1}, \ldots, e^{j \angle x^{\star}_N}]^T.
\end{equation}
The computation of \eqref{Eq: Closed form optimal RIS total MSE} is dominated by the eigenvalue decomposition of a $K \times K$ matrix and the inverse of the $K \times K$ matrix $\mathbf{Q}$. Hence, the computational complexity of \eqref{Eq: Closed form optimal RIS total MSE} is $\mathcal{O}(K^3)$.

In Summary, we can use the \eqref{Eq: Closed form optimal RIS total MSE} as an approximate solution to the minimization of $\mathrm{MSE}_{\mathrm{Tot}}$ \eqref{Eq: MSE metric}. In this paper, we refer to \eqref{Eq: Closed form optimal RIS total MSE} as the $\mathrm{MSE}_{\mathrm{Tot}}$ solution.

\section{Results}\label{Sec: Results}
We now demonstrate the effectiveness of the different techniques presented in Sec.~\ref{Sec: Optimization}. In the simulations, users were randomly located in a cell with a radius of 70m, outside exclusion radii of 5m around the BS and RIS. As stated in Sec.~\ref{Sec: Channel Model}, the steering vectors used in the channels are topology dependent. Here, we assume an $M$-element vertical uniform rectangular array (VURA) in the $y-z$ plane \cite{8812955}   with equal spacing in both dimensions at both the BS and RIS. The $y$ and $z$ components of a generic VURA steering vector at the BS for a given elevation angle, $\theta$, and azimuth angle, $\phi$, are given by,
\begin{align*}
    \mathbf{a}_{\mathrm{b},y}\left( \theta,\phi \right)
    &= 
    [1,\ldots,e^{j2\pi(M_y-1)d_{\mathrm{b}}\sin(\theta)\sin(\phi)} ]^T, \\
    \mathbf{a}_{\mathrm{b},z}\left( \theta,\phi \right)
    &= 
    [1,\ldots,e^{j2\pi(M_z-1)d_{\mathrm{b}}\cos(\theta)} ]^T,
\end{align*}
where $M = M_{y}M_{z}$ with $M_{y}, M_{z}$ denoting the number of antenna columns, rows respectively and  $d_{\mathrm{b}}=0.5$ is the antenna separation in wavelength units. Similarly at the RIS, we have
\begin{align*}
    \mathbf{a}_{\mathrm{r},y}\left( \theta,\phi \right)
    &= 
    [1,\ldots,e^{j2\pi(M_y-1)d_{\mathrm{r}}\sin(\theta)\sin(\phi)} ]^T, \\
    \mathbf{a}_{\mathrm{r},z}\left( \theta,\phi \right)
    &= 
    [1,\ldots,e^{j2\pi(M_z-1)d_{\mathrm{r}}\cos(\theta)} ]^T,
\end{align*}
where $N = N_{y}N_{z}$ with $N_{y},N_{z}$ denoting the number of columns, rows of RIS elements and $d_{\mathrm{r}}=0.2$ is the RIS element separation in wavelength units. The generic VURA steering vectors at the BS and RIS are then given by,
\begin{equation}\label{Eq: Steering vectors at BS and RIS}
\begin{split}
    \mathbf{a}_{\mathrm{b}}\left( \theta,\phi \right) &= \mathbf{a}_{\mathrm{b},y}\left( \theta,\phi \right) \otimes \mathbf{a}_{\mathrm{b},z}\left( \theta,\phi \right),
    \\
    \mathbf{a}_{\mathrm{r}}\left( \theta,\phi \right) &= \mathbf{a}_{\mathrm{r},y}\left( \theta,\phi \right) \otimes \mathbf{a}_{\mathrm{r},z}\left( \theta,\phi \right).
\end{split}
\end{equation}
Note that \eq{\ref{Eq: Steering vectors at BS and RIS}} can be used to generate all of the channels in Sec.~\ref{Sec: Channel Model} by substituting the relevant elevation and azimuth angles. 
For the LOS components in channels $\Hd{}$ and $\Hru{}$, the elevation and azimuth AOAs for the $\kth{k}$ UE are generated using $\theta_{\mathrm{d}}^{(k)}, \theta_{\mathrm{ru}}^{(k)}\sim \mathcal{U}[0,\pi]$, $\phi_{\mathrm{d}}^{(k)}, \phi_{\mathrm{ru}}^{(k)} \sim \mathcal{U}[-\pi/2,\pi/2]$. For the LOS component of $\myVM{H}{br}{}$, we assume that the elevation and azimuth angles are selected as follows, with less variation in elevation than azimuth: 
$ \theta_{D} \sim \mathcal{U}[70^{o},90^{o}] $, $ \phi_{D} \sim \mathcal{U}[-30^{o},30^{o}] $,
$ \theta_{A} = 180^{o} - \theta_{D} $, $ \phi_{A} \sim \mathcal{U}[-30^{o},30^{o}] $. 

For the rays in the scattered components, we model all central and deviation elevation angles by \cite{8812955}: $\theta_{\mathrm{E},c}^{(k)}\sim\mathcal{L}(1/\hat{\sigma}_{\mathrm{E},c}) , \delta_{\mathrm{E},c,s}^{(k)}\sim\mathcal{L}(1/\hat{\sigma}_{\mathrm{E},s})$, and the central and deviation azimuth angles by $\phi_{\mathrm{E},c}^{(k)}\sim\mathcal{N}(\mu_{\mathrm{E},c},\sigma_{\mathrm{E},{c}}^2),  \Delta_{\mathrm{E},c,s}^{(k)}\sim\mathcal{L}(1/\sigma_{\mathrm{E},s})$. The subscript $\mathrm{E} \in \{\mathrm{d},\mathrm{ru},\mathrm{br}\}$ represents the different channels. For both $\Hd{}$ and $\Hru{}$, we assume that the rays are broadly spread with identical parameter values for generating the subrays for each cluster in both channels. For channel $\Hbr{}$, we assume that the rays are narrowly spread, for which the parameter values are also given in \cite{8812955}. All system parameter values are given in Table \ref{Table 1} and remain unchanged unless otherwise specified. Note that the parameter values for the path loss exponents and distances related to the deployment of the BS, RIS and UEs are adapted from \cite{ParameterValues}.
\begin{table}[h]
    \centering
    \begin{tabular}{|c | c  |} 
     \hline
     Parameter & Values \\
     \hline
     Cell Radius & 70 m \\
     Exclusion Radius & 5 m \\
     BS Antennas, $M$ & 32 \\
     Path Loss at 1m, $P$ & -30 dB \\
     Path Loss Exponent, $\gamma_{\mathrm{ru}}, \gamma_{\mathrm{d}}$ & 2.8, 3.5 \\
     Noise Power, $\sigma^2$ & -80 dBm \\
     RIS-BS Distance, $d_{\mathrm{br}}$ & 51 m \\
     \hline
     \underline{Channels $\Hd{}, \Hru{}$} &\\
     $C_{\mathrm{d}}=C_{\mathrm{ru}}$ & 20 \\
     $S_{\mathrm{d}}=S_{\mathrm{ru}}$ & 20 \\
     $\mu_{\mathrm{d},c}=\mu_{\mathrm{ru},c}$, & 0$^{\circ}$ \\
     $\sigma^2_{\mathrm{d},c}=\sigma^2_{\mathrm{ru},c}, 
     \sigma^2_{\mathrm{d},s}=\sigma^2_{\mathrm{ru},s}$ & 31.64$^{\circ}$, 24.25$^{\circ}$ \\
     $\hat{\sigma}^2_{\mathrm{d},c}=\hat{\sigma}^2_{\mathrm{ru},c},\hat{\sigma}^2_{\mathrm{d},s}=\hat{\sigma}^2_{\mathrm{ru},s}$ & 6.12$^{\circ}$, 1.84$^{\circ}$ \\
     \hline 
     \underline{Channel $\Hbr{}$} & \\
     $C_{\mathrm{br}}$, $S_{\mathrm{br}}$ & 3, 16 \\
     $\mu_{\mathrm{br},c}$ & 0$^{\circ}$ \\
     $\sigma^2_{\mathrm{br},c},
     \sigma^2_{\mathrm{br},s},\hat{\sigma}^2_{\mathrm{br},c},\hat{\sigma}^2_{\mathrm{br},s}$ & 14.4$^{\circ}$, 6.24$^{\circ}$, 1.9$^{\circ}$, 1.37$^{\circ}$ \\
     \hline 
    \end{tabular}
    \caption{System parameter values}
    \label{Table 1}
\end{table}

\begin{figure}[h]
	\centering
	\myincludegraphics{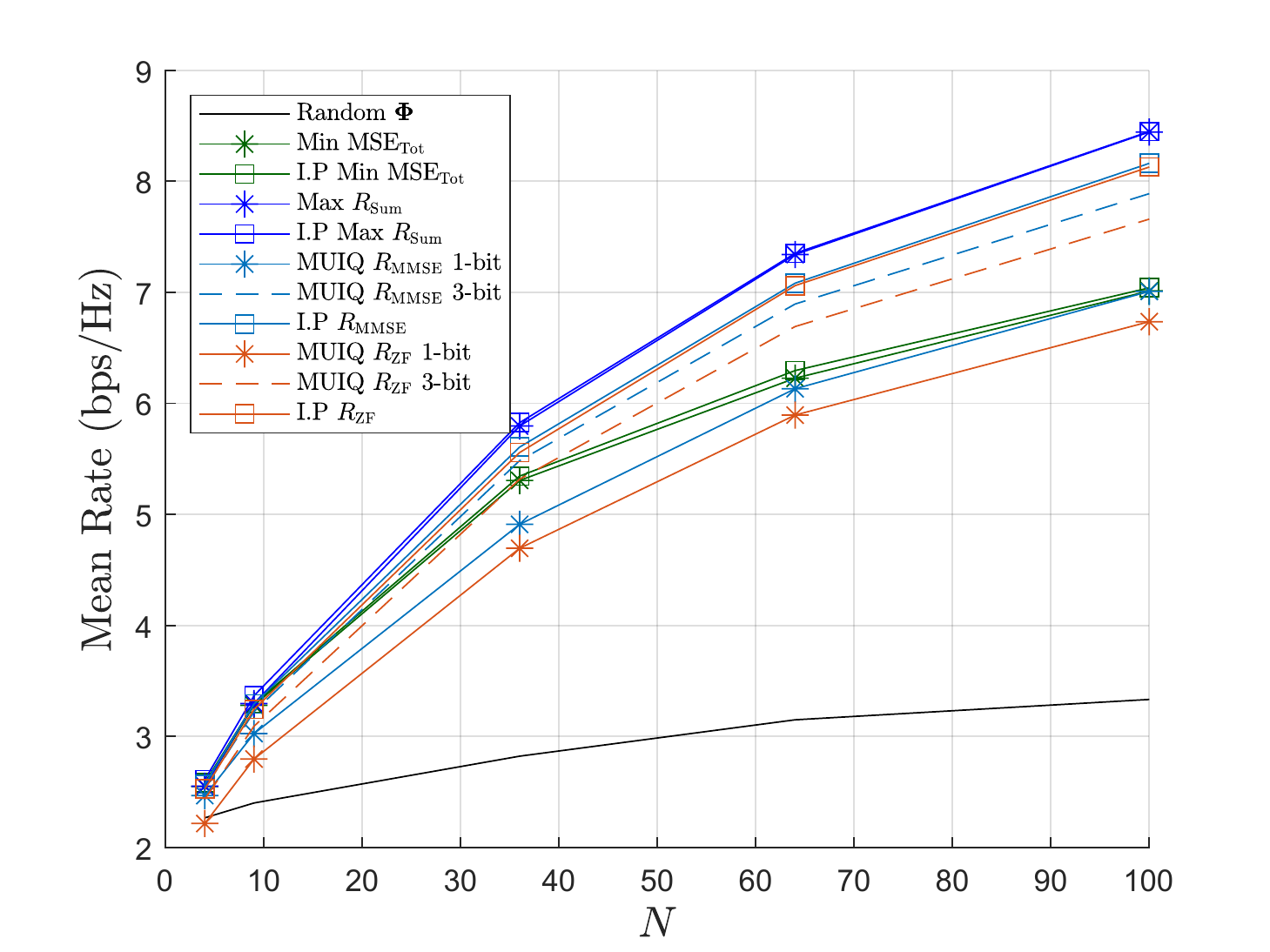}
	\caption{Average sum-rate metrics for varying $N$ and $\kru=\kd=1,\kbr=\infty$, $ K=2 $, $L=1$.}
	\label{Fig: Fig 1 K=2 kbr=inf}
\end{figure}

\begin{figure}[h]
	\centering
	\myincludegraphics{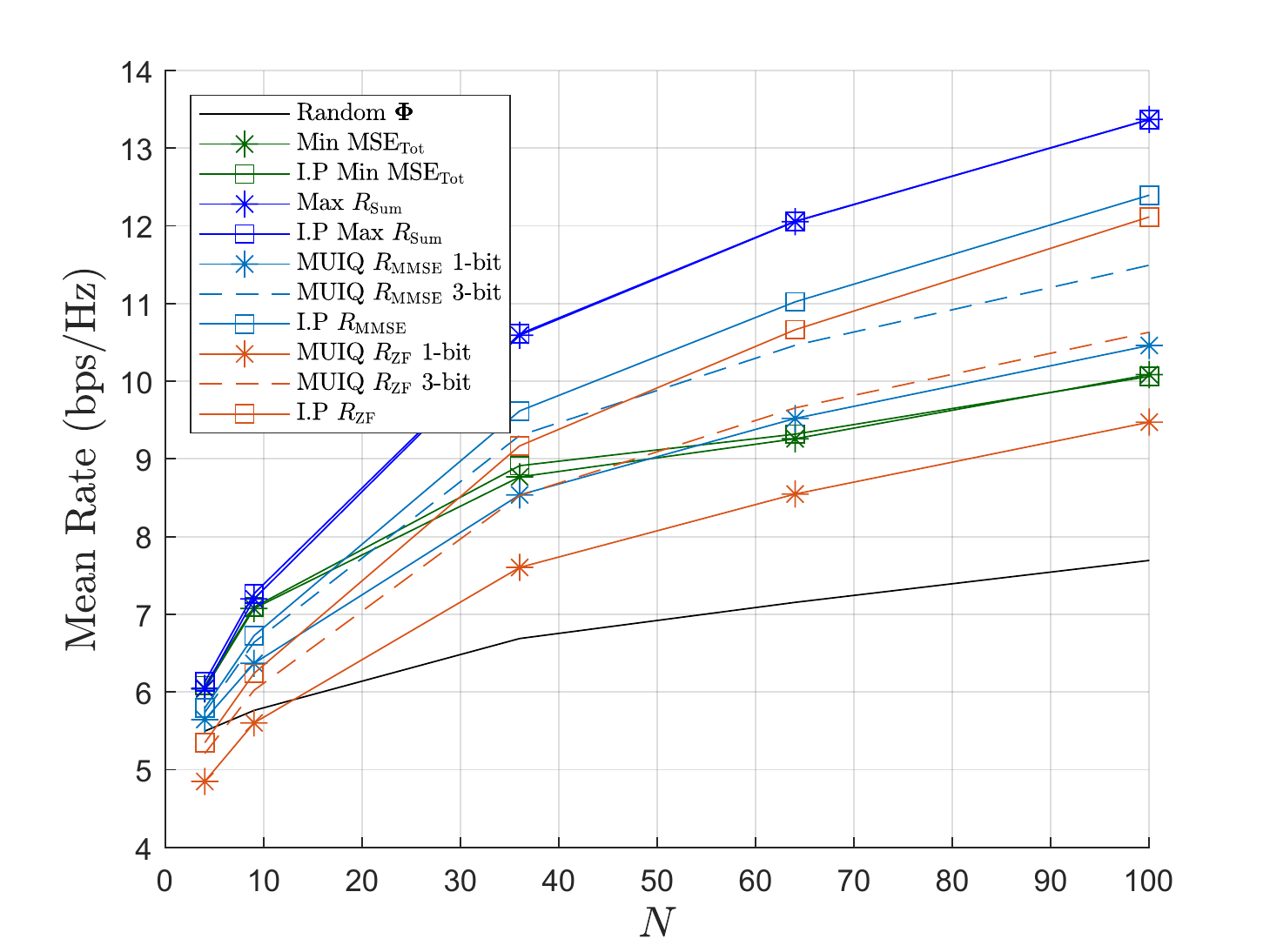}
	\caption{Average sum-rate metrics for varying $N$ and $\kru=\kd=1,\kbr=\infty$, $ K=5$, $L=1$.}
	\label{Fig: Fig 1 K=5 kbr=inf}
\end{figure}

In Fig.~\ref{Fig: Fig 1 K=2 kbr=inf} and Fig.~\ref{Fig: Fig 1 K=5 kbr=inf}, we demonstrate the effectiveness of the optimization techniques presented in Sec.~\ref{Sec: Optimization} for varying RIS sizes and UE numbers. 
Here, we show the sum rate when the RIS matrix is set to the sum rate solution given by \eqref{Eq: Closed form optimal RIS}  and also when using the $\mathrm{MSE}_{\mathrm{Tot}}$ solution given by \eqref{Eq: Closed form optimal RIS total MSE}. 
These results represent the use of closed form RIS phase solutions to optimize system performance for continuous RIS phases. For discrete RIS phases, Fig.~\ref{Fig: Fig 1 K=2 kbr=inf} and Fig.~\ref{Fig: Fig 1 K=5 kbr=inf} also show the results of using Algorithm~\ref{Algorithm: MUIQ} to maximize $R_{\mathrm{ZF}}$ and $R_{\mathrm{MMSE}}$ for $b \in \{1,3\}$ bits and $L=1$ iterations. All of these expressions are computed for scenarios where $\kru=\kd=1$ and $\kbr=\infty$ to represent user channels containing both scattered and LOS components 
and pure LOS RIS-BS channels, respectively. The number of UEs is $ K \in\{2,5\}$. The average sum rate results for each of the optimization techniques are compared to two benchmark cases:
\begin{itemize}
    \item the optimal sum-rate computed by built-in numerical optimization software using the interior point (I.P) algorithm; 
    \item the sum-rate  achieved by a random set of RIS phases selected from $\mathcal{U}[0,2\pi]$. 
\end{itemize}
As the results of several algorithms are shown in the figures, for clarity we also define the methodology associated with each figure legend in Table~\ref{Table figs}.

\begin{table}[h]
    \centering
    \begin{tabular}{|c | c  |} 
     \hline
     Legend entry & RIS design algorithm \\
     \hline
     Random & Elements of $\Phi$ are i.i.d. $\mathcal{U}[0,2\pi]$ \\
     Min $\mathrm{MSE}_{\mathrm{Tot}}$& Elements of $\Phi$ designed using \eqref{Eq: Closed form optimal RIS total MSE} \\
     I.P Min $\mathrm{MSE}_{\mathrm{Tot}}$ &  I.P algorithm to minimize $\mathrm{MSE}_{\mathrm{Tot}}$ \\
     Max ${R}_{\mathrm{Sum}}$ &  Elements of $\Phi$ designed using \eqref{Eq: Closed form optimal RIS} \\
     I.P Max ${R}_{\mathrm{Sum}}$ &  I.P algorithm to maximize ${R}_{\mathrm{Sum}}$ \\
    MUIQ ${R}_{\mathrm{MMSE}}$ & Algorithm 1 applied to ${R}_{\mathrm{MMSE}}$  \\
    I.P  ${R}_{\mathrm{MMSE}}$ &  I.P algorithm to maximize ${R}_{\mathrm{MMSE}}$ \\
       MUIQ  ${R}_{\mathrm{ZF}}$ &  Algorithm 1 applied to ${R}_{\mathrm{ZF}}$\\
    I.P ${R}_{\mathrm{ZF}}$ &  I.P algorithm to maximize ${R}_{\mathrm{ZF}}$ \\
    
     \hline 
    \end{tabular}
    \caption{RIS design methods used in Figs.~\ref{Fig: Fig 1 K=2 kbr=inf}-\ref{Fig: Fig 1 K=5 kbr=inf, MURIQ}}
    \label{Table figs}
\end{table}

For continuous RIS phases, the use of \eqref{Eq: Closed form optimal RIS} to maximize $R_{\mathrm{sum}}$ and \eqref{Eq: Closed form optimal RIS total MSE} to minimize the total mean square error achieves results that are extremely close to those obtained using the interior point method. Hence, the closed form solutions given by \eqref{Eq: Closed form optimal RIS} and \eqref{Eq: Closed form optimal RIS total MSE} are  highly effective in maximizing $R_{\mathrm{sum}}$ and minimizing $\mathrm{MSE}_{\mathrm{Tot}}$, respectively. However, notice that as the number of RIS elements increase, the sum rates produced by minimizing $\mathrm{MSE}_{\mathrm{Tot}}$ deviate from the maximization of $R_{\mathrm{ZF}}$ and $R_{\mathrm{MMSE}}$. This is the trade-off for the low complexity design based on $\mathrm{MSE}_{\mathrm{Tot}}$ and channel separation.

\begin{figure}[h]
	\centering
	\myincludegraphics{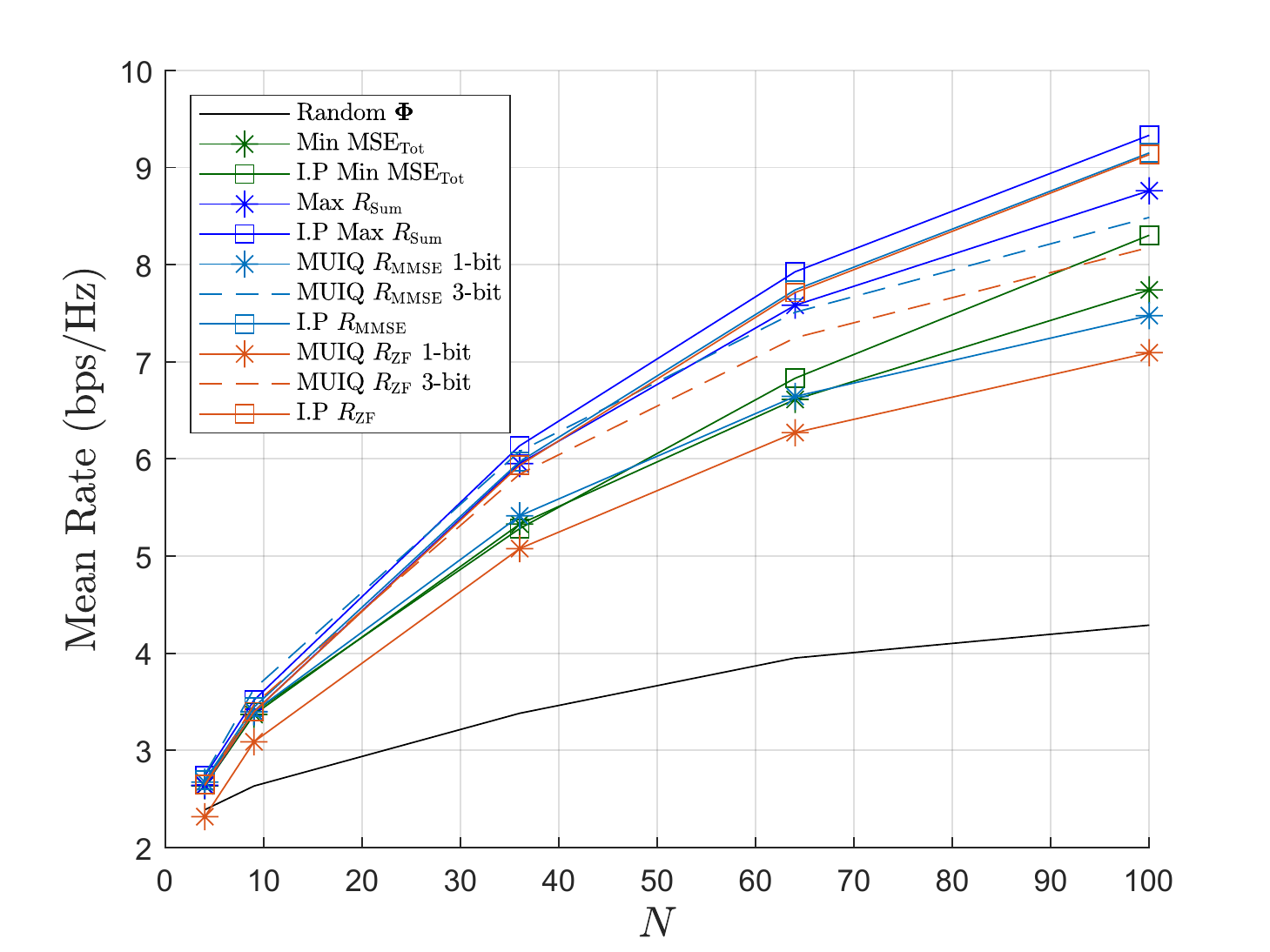}
	\caption{Average sum-rate metrics for varying $N$ and $\kru=\kd=1,\kbr=1$, $ K=2 $, $L=1$.}
	\label{Fig: Fig 1 K=2 kbr=1}
\end{figure}

\begin{figure}[h]
	\centering
	\myincludegraphics{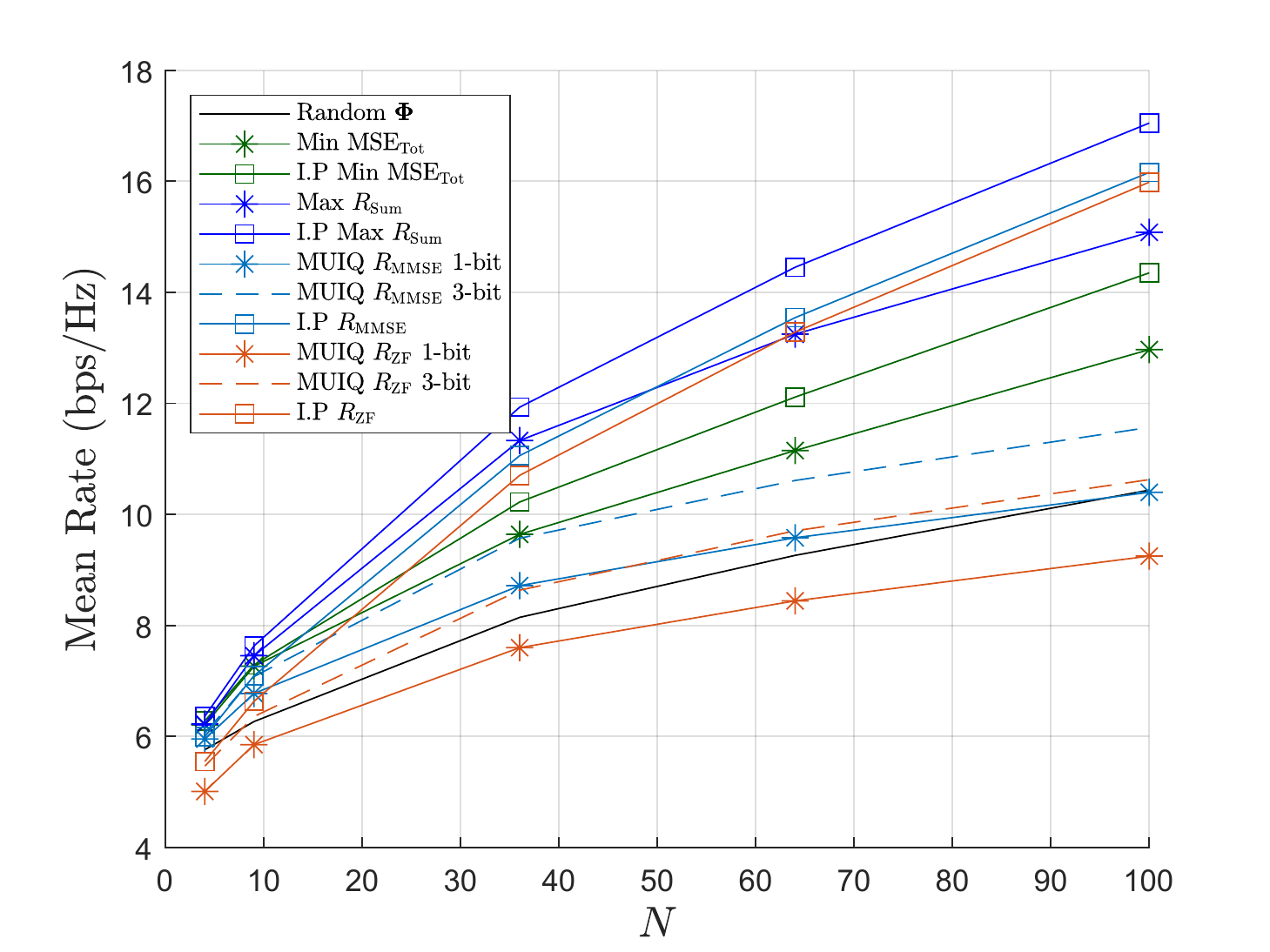}
	\caption{Average sum-rate metrics for varying $N$ and $\kru=\kd=1,\kbr=1$, $ K=5$, $L=1$.}
	\label{Fig: Fig 1 K=5 kbr=1}
\end{figure}

Note that these observations are for scenarios where the RIS-BS channel is LOS ($\kbr =\infty$). Since the optimization techniques in Sec.~\ref{Sec: Optimization} are designed for a system where the RIS-BS channel is only LOS, it is worth investigating the robustness of these optimization techniques to scattered RIS-BS channels. This is done in Fig.~\ref{Fig: Fig 1 K=2 kbr=1} and Fig.~\ref{Fig: Fig 1 K=5 kbr=1} where all system parameters remain unchanged except for $\kbr=1$ which represents equal scattered and LOS powers in the RIS-BS channel. Note that equal scattered and LOS power is  very different to the pure LOS assumption on which the design was based. Hence, this is a challenging test of robustness.
From Fig.~\ref{Fig: Fig 1 K=2 kbr=1} and Fig.~\ref{Fig: Fig 1 K=5 kbr=1}, notice that the sum rate solution \eqref{Eq: Closed form optimal RIS} and the $\mathrm{MSE}_{\mathrm{Tot}}$ solution \eqref{Eq: Closed form optimal RIS total MSE} achieve similar rates to the results obtained through the interior point algorithm. Hence, even with  channels with a strong scattered component, these closed form solutions for the RIS matrix achieve  useful sum rate results. For example, both Fig.~\ref{Fig: Fig 1 K=2 kbr=1} and Fig.~\ref{Fig: Fig 1 K=5 kbr=1} show that around 90\% of the optimal  ${R}_{\mathrm{Sum}}$ value is achieved.

The difference between rates from the interior point algorithm and the optimization techniques in Sec.~\ref{Sec: Optimization} become more prominent when the number of UEs increases. Fig.~\ref{Fig: Fig 1 K=2 kbr=1} and Fig.~\ref{Fig: Fig 1 K=5 kbr=1} show the results for $K=2$ and $K=5$ UEs, respectively. Comparing Fig.~\ref{Fig: Fig 1 K=5 kbr=1} and Fig.~\ref{Fig: Fig 1 K=5 kbr=inf}, we note that the separation gap between Algorithm~\ref{Algorithm: MUIQ} and the interior point method for both $R_{\mathrm{ZF}}$ and $R_{\mathrm{MMSE}}$ increases when the RIS-BS channel becomes more scattered. We also note that in Fig.~\ref{Fig: Fig 1 K=5 kbr=1}, a random RIS matrix is capable of achieving rates  better than the MUIQ Algorithm~\ref{Algorithm: MUIQ} with low level quantization. Hence, when the LOS assumption is relaxed, a higher level of quantization is required. Nevertheless, 3-bit MUIQ gives $R_{\mathrm{MMSE}}$ values around 70\% and 90\% of optimum for $K=2$ and $K=5$ respectively. This is very promising considering the very large difference between the RIS-BS channel used and the channel assumed for design.

The results produced by Algorithm~\ref{Algorithm: MUIQ} thus far are for a single iteration (i.e. $L=1$). We now investigate the effects of more iterations with $L=2$, which are shown in Fig.~\ref{Fig: Fig 1 K=2 kbr=inf, MURIQ} and Fig.~\ref{Fig: Fig 1 K=5 kbr=inf, MURIQ} for the case of a pure LOS RIS-BS channel.
\begin{figure}[h]
	\centering
	\myincludegraphics{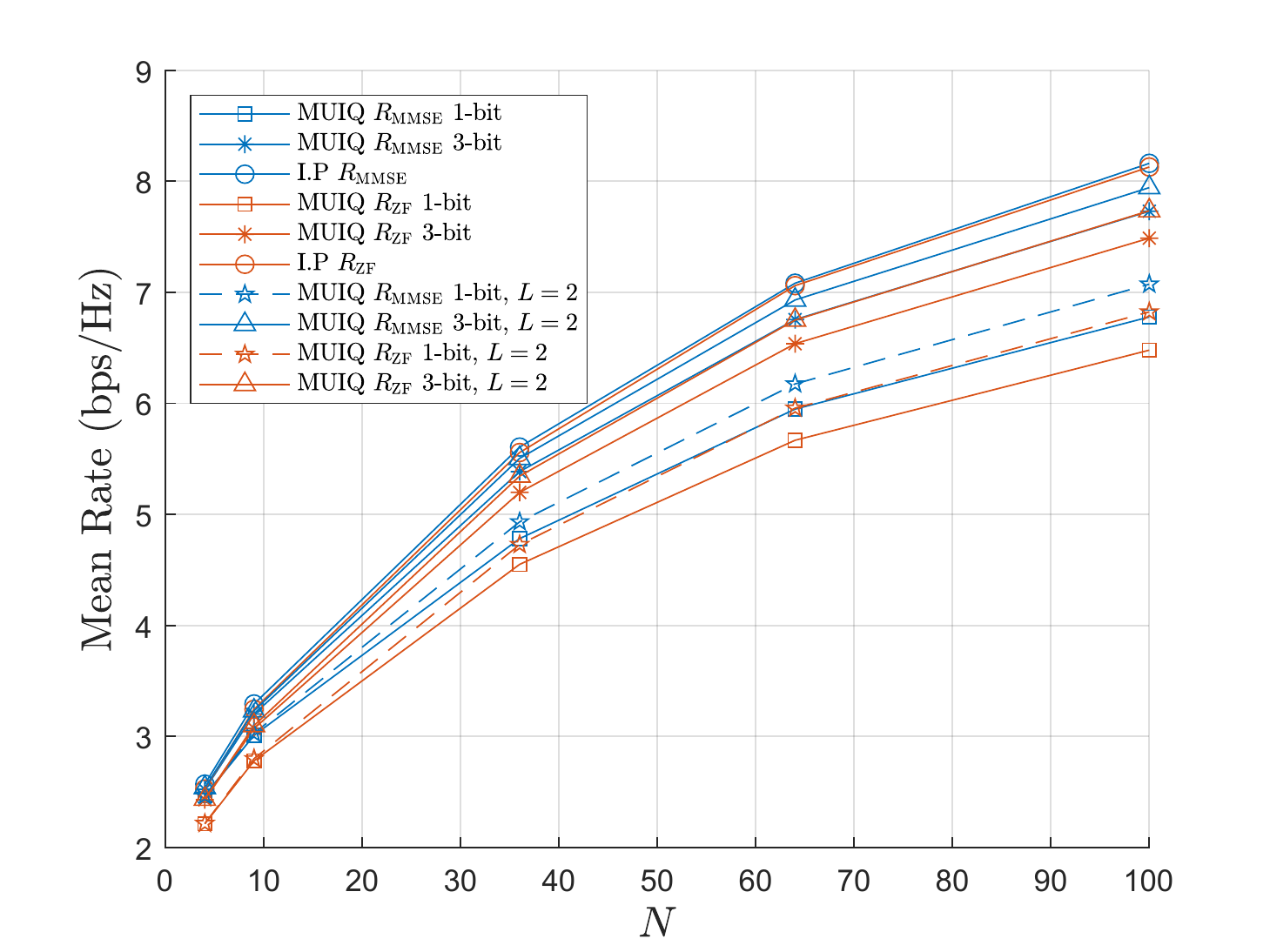}
	\caption{Average sum-rate metrics for varying $N$ and $\kru=\kd=1,\kbr=\infty$, $ K=2 $, $L=2$.}
	\label{Fig: Fig 1 K=2 kbr=inf, MURIQ}
\end{figure}
\begin{figure}[h]
	\centering
	\myincludegraphics{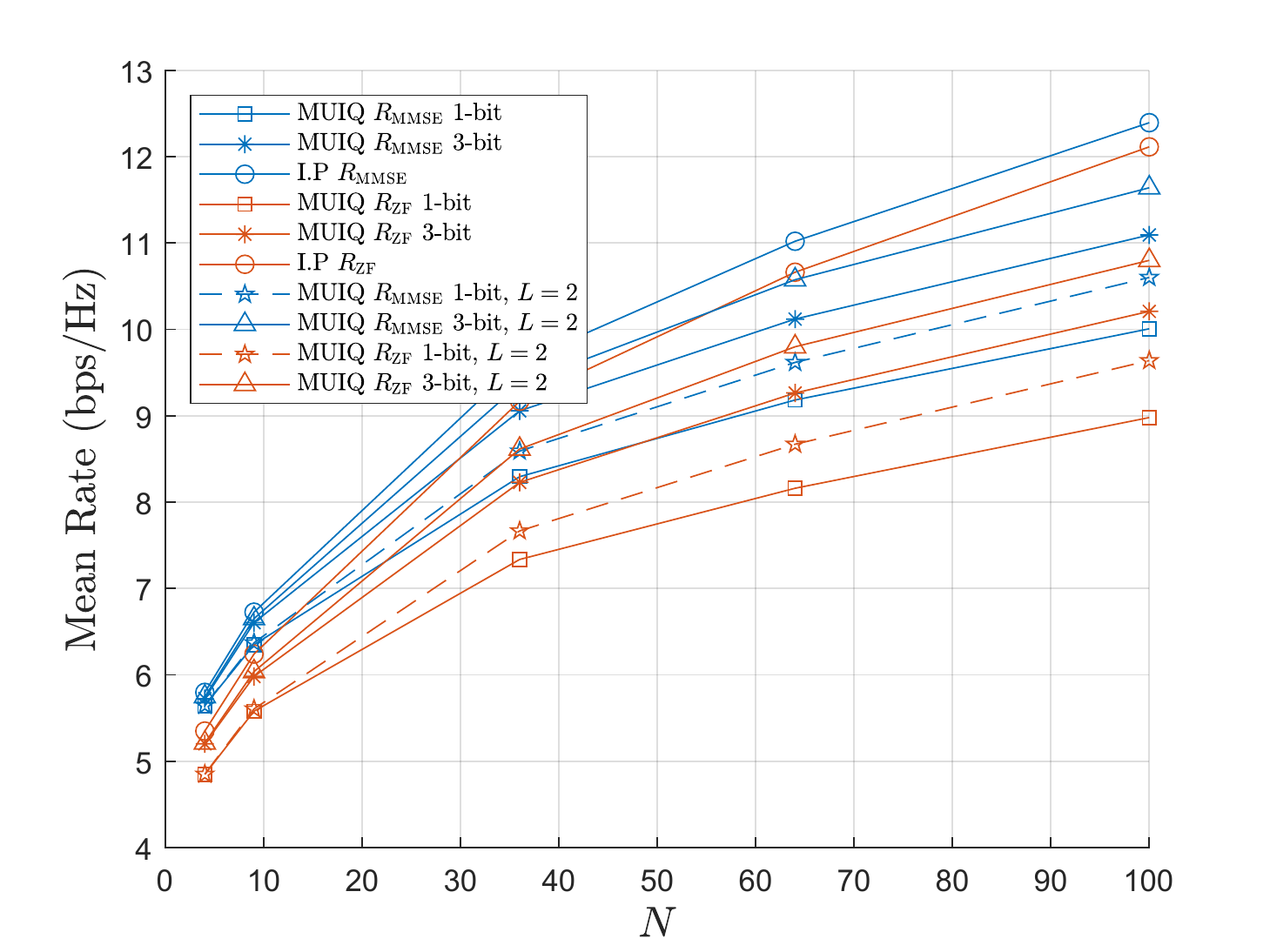}
	\caption{Average sum-rate metrics for varying $N$ and $\kru=\kd=1,\kbr=\infty$, $ K=5$, $L=2$.}
	\label{Fig: Fig 1 K=5 kbr=inf, MURIQ}
\end{figure}
Observe that by increasing the number of iterations in Algorithm~\ref{Algorithm: MUIQ}, the sum rates for ZF and MMSE linear receivers improve and approach the sum rate results of the interior point algorithm. Note that the improvement in sum rates due to increased iterations are most noticeable for large RIS sizes but the improvement over $L=1$ is only a few percent. This supports the use of Algorithm 1 as a low complexity approach, particularly for the important case when $b$ is small and only one iteration is employed. 

In summary, when the RIS-BS channel is LOS, Fig.~\ref{Fig: Fig 1 K=2 kbr=inf} and Fig.~\ref{Fig: Fig 1 K=5 kbr=inf} show that the optimization techniques developed in Sec.~\ref{Sec: Optimization} perform extremely well for discrete and continuous RIS phases and achieve large fractions of the optimum rates, even with low bit quantization. Introducing large amounts of scattering into the RIS-BS channel, it is observed in Fig.~\ref{Fig: Fig 1 K=2 kbr=1} and Fig.~\ref{Fig: Fig 1 K=5 kbr=1} that the designs are fairly robust to this deviation from the design assumption. Finally, in Fig.~\ref{Fig: Fig 1 K=2 kbr=inf, MURIQ} and Fig.~\ref{Fig: Fig 1 K=5 kbr=inf, MURIQ} we show that multiple iterations of Algorithm~\ref{Algorithm: MUIQ} improves performance, but $L=1$ remains a high-performance, low complexity solution.

\section{Conclusion}
In this paper, we have developed a novel \textit{channel separation} technique which allows for a better understanding of the effects of the RIS phases on the sum rate performance. Specifically, chanel separation creates an equivalent channel matrix separated into two parts; one independent of the RIS and another part consisting of a single row directly impacted by the RIS. Leveraging channel separation, we propose a simple iterative algorithm to maximize the sum rates of ZF and MMSE linear receivers for discrete RIS phases with $b-$level quantization. For continuous RIS phases, we present closed form RIS phase expressions to maximize the traditional sum-rate and to minimize the total mean square error metrics. The latter metric is presented as an alternative to maximizing sum rates for ZF and MMSE linear receivers. Numerical results demonstrate the effectiveness of the optimization techniques. For discrete RIS phases, the proposed algorithm is capable of achieving sum rates close to those obtained through a full numerical interior point optimization procedure, even with low level RIS quantization. Increasing the number of iterations of the algorithm improves the sum rate. For continuous RIS phases, our closed form phase solutions achieve sum rates very close to those for numerical optimization. When the RIS-BS channel becomes scattered, the proposed algorithm for discrete RIS phases weakens as channel separation was designed for systems where the RIS-BS link is LOS. However, even with scattered RIS-BS channels, the closed form solutions for continuous RIS phases are robust.

\begin{appendices}
\section{Maximum Eigenvector Method}\label{App: finding max eigenvector}
 Let $\mathbf{Y} = \alpha\mathbf{I}_K + \myVM{B}{}{}\myVM{C}{}{H}\myVM{C}{}{}$ where $\alpha$ is a positive constant, $\myVM{C}{}{}$ is an $N \times K$ matrix, $\myVM{B}{}{}$ is a $K \times K$ Hermitian matrix and $K<N$. Let $\myVM{y}{}{}$ be the maximum eigenvector of $\myVM{Y}{}{}$ with eigenvalue $\lambda_1$, then
 \begin{equation}\notag
(\alpha\mathbf{I}_K + \myVM{B}{}{}\myVM{C}{}{H}\myVM{C}{}{})\myVM{y}{}{}=\lambda_1\myVM{y}{}{}.
\end{equation}
Multiplying by $\myVM{C}{}{}$ gives
 \begin{equation}\notag
(\alpha \myVM{C}{}{} + \myVM{C}{}{}\myVM{B}{}{}\myVM{C}{}{H}\myVM{C}{}{})\myVM{y}{}{}=\lambda_1\myVM{C}{}{}\myVM{y}{}{}.
\end{equation}
Defining $\myVM{x}{}{}=\myVM{C}{}{}\myVM{y}{}{}$ gives
 \begin{equation}\notag
(\alpha \mathbf{I}_N + \myVM{C}{}{}\myVM{B}{}{}\myVM{C}{}{H})\myVM{x}{}{}=\lambda_1\myVM{x}{}{}.
\end{equation}
Hence, $\myVM{C}{}{}\myVM{y}{}{}$ is an eigenvector of $\alpha \mathbf{I}_N + \myVM{C}{}{}\myVM{B}{}{}\myVM{C}{}{H}$ and it is the maximum eigenvector because the eigenvalues of $ \myVM{B}{}{}\myVM{C}{}{H}\myVM{C}{}{}$ are equal to the non-zero eigenvalues of $\myVM{C}{}{}\myVM{B}{}{}\myVM{C}{}{H}$.

\end{appendices}

\bibliographystyle{IEEEtran}
\bibliography{MURIQ_Paper}
\end{document}